\documentclass[12pt]{iopart}
\usepackage{graphicx}
\usepackage{dcolumn}
\usepackage{bm}

\newcommand{\ic}{\mathrm{i}}
\newcommand{\ee}{\mathrm{e}}
\newcommand{\ud}{\mathrm{d}}
\newcommand{\erf}{\mathrm{erf}}

\newcommand{\im}{\mathrm{Im }}
\newcommand{\re}{\mathrm{Re }}
\newcommand{\rR}{\mathbb{R}}

\expandafter\let\csname equation*\endcsname\relax 
\expandafter\let\csname endequation*\endcsname\relax 

\usepackage{amsmath}
\usepackage{amssymb}
\usepackage{color}
\usepackage{cite,soul}
\newcommand{\old}[1]{}

\begin{document}

\title{Bohmian trajectories for the half-line barrier}

\author{R\'emy Dubertrand$^{1}$\footnote{remy.dubertrand@uliege.be}, Jeong-Bo Shim$^{1}$\footnote{jbshim@uliege.be} and Ward Struyve$^{2}$\footnote{ward.struyve@gmail.com}}

\address{
$^1$  Institut de Physique Nucl\'eaire, Atomique et de Spectroscopie,
CESAM, Universit\'e de Li\`ege, B\^at. B15, B - 4000 Li\`ege, Belgium\\
$^{2}$Mathematisches Institut, Ludwig-Maximilians-Universit\"at M\"unchen, Theresienstr.\ 39, 80333 M\"unchen, Germany}

\maketitle  

\begin{abstract}
\noindent
Bohmian trajectories are considered for a particle that is free (i.e.\ the potential energy is zero), except for a half-line barrier. On the barrier, both Dirichlet and Neumann boundary conditions are considered. The half-line barrier yields one of the simplest cases of diffraction. Using the exact time-dependent propagator found by Schulman, the trajectories are {computed numerically} for different initial Gaussian wave packets. In particular, it is found that different boundary condition{s} may lead to qualitatively different sets of trajectories. In the Dirichlet case, the particles tend to be more strongly repelled. The case of an incoming plane wave is also considered. The corresponding Bohmian trajectories are compared with the trajectories of an oil drop hopping {on} the surface of a vibrating bath.
\end{abstract}

\section{Introduction}

Bohmian mechanics (also known as de\ Broglie-Bohm theory or pilot-wave theory) describes point-particles moving in physical space with a velocity that depends on the wave function (which satisfies Schr\"odinger's equation) \cite{bohm93,holland93b,duerr09}. It is an alternative to standard quantum theory that is free of the conceptual problems, such as the measurement problem, that plague the latter. It reproduces the predictions of standard quantum theory insofar the latter are unambiguous. 
{Bohmian mechanics has been used for more practical purposes recently: electronic transport \cite{Albareda2013}, molecular reaction \cite{Curchod2011}, see e.g. the recent review \cite{benseny14}.}
Another virtue of Bohmian mechanics is that it allows for the visualisation of physical processes, in particular diffraction situations. Many examples have been presented in the literature: single slit, double slit, diffraction by gratings, see e.g.\ \cite{philippidis79,gondran05,sanz08,sanz14,sanz15}. Recently, using weak measurements, the Bohmian trajectories have also been found experimentally in the case of the double-slit experiment for a single particle \cite{kocsis11} and for entangled particles \cite{mahler16}. Also recently, the Bohmian dynamics is being studied in the context of hydrodynamic analogues of quantum mechanics \cite{couder06,andersen15,bush15,dubertrand16}. These analogues concern silicon oil droplets bouncing on the surface of a harmonically vibrating bath. The bounces of the droplets, dubbed {\em walkers}, create surface waves which in turn influences the droplet dynamics. These systems were claimed to show some analogy with Bohmian mechanics due to the coupling between droplet and wave.

In diffraction situations it is often hard to find solutions to the Schr\"odinger equation. Often simplifications are introduced. A general solution can be expressed in terms of the quantum time propagator. In the case of a single slit, the propagator is only expressible as a Fourier integral of a series of special functions (Mathieu functions) \cite{hsu60}. In the case of two slits the propagator is not known and often approximations are used, see e.g.\ \cite{beau12}. In the first simulation of the Bohmian trajectories for the double-slit experiment, a superposition of Gaussian wave packets is used to model the actual slits \cite{philippidis79}. A much more realistic yet approximate modelling was done in \cite{gondran05}, where the wave function was numerically integrated.

In the present paper, we consider the half-line barrier. For this system, the quantum mechanical propagator is exactly known and is given in terms of special functions (Fresnel integrals) \cite{schulman82}. This greatly simplifies the numerical simulations for the Bohmian trajectories. We will consider both Neumann and Dirichlet boundary conditions, and \old{we will}present the trajectories in the case of initial Gaussian wave functions and stationary states with different momenta. 
{While Neumann boundary conditions do not seem relevant for quantum systems, they are very common in fluid dynamics, especially relevant here for a direct comparison with the droplet trajectory. Indeed Faraday surface waves obey those boundary conditions in the limit of vanishing viscosity \cite{ursell1954}.}
As such we provide a visualisation of one of the simplest quantum diffraction situations. It is also of interest for comparing with the droplet dynamics. The reason is that it is still unclear to what extent the Bohmian dynamics is similar to droplet dynamics. While the initial work of Couder and Fort \cite{couder06} showed a great similarity with quantum mechanics concerning the droplet distribution in the case of a double slit experiment, more recent attempts to reproduce this result have so far not been successful \cite{harris15,andersen15,batelaan16}. It is therefore important to investigate the possible similarities in a much more simpler system like in the case of diffraction from a half-line barrier. Such an investigation has been initiated in \cite{harris15}.

The outline of the paper is as follows. First, in section \ref{bd} we give an introduction to Bohmian mechanics and some details concerning the propagator method for solutions to the Schr\"odinger equation. Then in section \ref{half-plane} we consider the case of a Bohmian particle in the half-plane, which corresponds to having an infinite potential barrier along a line. We turn to the half-line barrier in section \ref{half-line}. For the half-line barrier we consider Gaussian initial wave packets and stationary states. The latter case is important for the comparison with the droplet dynamics. Our conclusions are presented in section \ref{conclusion}. 

\section{Bohmian dynamics}\label{bd}
The Bohmian dynamics for {a} single particle with position ${\bf X}$ is given by 
\begin{equation}
   \label{Bohm_eq}
{\dot {\bf X}}(t)  =  {\bf v}^\psi({\bf X}(t){,t})\ ,
\end{equation}
where 
\begin{equation}
  {\bf v}^\psi({\bf x},t) =  \frac{\hbar}{m} \im \frac{{\boldsymbol \nabla} \psi( {\bf x},t)}{\psi( {\bf x},t)} = \frac{1}{m}{\boldsymbol \nabla} S( {\bf x},t)\ ,\label{defvpsi}
\end{equation}
with $m$ the mass and $\psi=|\psi|\ee^{\ic S/\hbar}$ the wave function which satisfies the Schr\"odinger equation
\begin{equation}
  \label{schro}
 \ic \hbar \frac{\partial}{\partial t}\psi( {\bf x},t) =  -\frac{\hbar^2}{2m} \nabla^2 \psi( {\bf x},t)+V( {\bf x}) \psi( {\bf x},t)\ .
\end{equation}
The Bohmian dynamics preserves the distribution $|\psi|^2$. That is, if the initial particle distribution is given by $|\psi({\bf x},t_0)|^2$ at some time $t_0$, then the distribution is given by $|\psi({\bf x},t)|^2$ at other times $t$. This property is called {\em equivariance} and is crucial in deriving the usual quantum predictions \cite{duerr09}. It follows from the fact that an arbitrary distribution $\rho({\bf x},t)$ of particles satisfies the continuity equation
\begin{equation}
 \frac{\partial \rho}{\partial t}  + {\boldsymbol \nabla}  \cdot ({\bf v}^\psi \rho) = 0\ ,
\end{equation}
which is also satisfied by $|\psi|^2$ as a consequence of the Schr\"odinger equation.

For the systems we will consider here, the effect of the potentials can be completely expressed in terms of Dirichlet or Neumann boundary conditions for the wave function. In the case of Dirichlet boundary conditions the wave function is zero along the boundary and as such the Bohmian velocity is not defined there. In the case of Neumann boundary conditions the directional derivative of the wave function normal to the boundary is zero and as such the component of the Bohmian velocity normal to the boundary vanishes.

We will often use the propagator to express solutions of the Schr\"odinger equation. For a single particle, solutions of the Schr\"odinger equation (\ref{schro}) will be written as
\begin{equation}
  \label{GWP_2D}
  \psi({\bf x},t)=\int K({\bf x},t|{\bf x}_0,0)  \psi_0({\bf x}_0)\ud {\bf x}_0\ ,
\end{equation}
for $t>0$ and where $\psi_0$ is the initial wave function at time $t=0$ and $K$ is the quantum propagator, which satisfies
\begin{align}
  \ic\hbar \dfrac{\partial }{\partial t}K({\bf x},t|{\bf x}_0,0) &= -\dfrac{\hbar^2}{2m} \nabla^2 K({\bf x},t|{\bf x}_0,0)\ ,{t>0}\\ 
\lim_{t\to 0^+} K({\bf x},t|{\bf x}_0,0)& =\delta({\bf x}-{\bf x}_0)\ ,& 
\end{align}
together with appropriate boundary conditions. In the case of Dirichlet boundary conditions, the propagator should be zero along the boundary, i.e.,
\begin{equation}
  \label{DirichletBC}
K_D({\bf x},t|{\bf x}_0,0)=0\ .
\end{equation}
In the case of Neumann boundary conditions, the normal derivative should be zero along the boundary, i.e.,
\begin{equation}
  \label{NeumannBC}
\dfrac{\partial }{\partial n}K_N({\bf x},t|{\bf x}_0,0)=0\ ,
\end{equation}
with $\partial/\partial n$ the normal derivative.  

A particular solution is given by the propagator itself, by taking
\begin{equation}
  \psi_0({\bf x})= \delta({\bf x}-{\bf x}_0)\ .
  \label{delta}
\end{equation}
While this solution is not in the Hilbert space, we can nevertheless consider the Bohmian trajectories. Namely, for the trajectories to be well defined, the wave function merely needs to be sufficiently differentiable. (For example, while plane waves are also not in the Hilbert space, one can consider the Bohmian trajectories, which correspond to straight lines in the direction of the wave vector.) In the case of the half-line barrier the trajectories corresponding with the propagator will be qualitatively very similar to those corresponding to an initially Gaussian wave function with zero momentum.

Throughout this paper, the units for all the plots are chosen such that $\hbar=1$ and $m=1/2$.

\section{Bohmian particle in the presence of a wall}\label{half-plane}

Consider a particle moving in the presence of a wall which restricts the motion to a half-plane. This problem is separable if the axis of Cartesian coordinates are conveniently chosen. In this section we will assume that the wall lies along the axis $y=0$. Then the problem can be split into two different one$-$dimensional problems. In the $x-$direction, it simply amounts to a free motion and the solution is recalled in \ref{Bohmianfree}. In the $y-$direction, the motion is restricted to the half-line $y>0$. 
  The wave functions we consider will be separable because both the initial wave function and the propagator are separable. We will treat the two one$-$dimensional problems separately before showing pictures for the full two$-$dimensional problem.

\subsection{Bohmian dynamics in one dimension: One particle in the presence of a wall}

In this section the dynamics is restricted to one Cartesian coordinate, which is denoted by $y$. In order to model a wall at $y=0$, it is first assumed that the wave function obeys Neumann boundary conditions at this point:
\begin{equation}
 { \frac{\partial \psi}{\partial y}\bigg|_{y=0}=0 \ .}
\end{equation}
Initially the wave packet, defined only along the half line $y\ge 0$, is a superposition of a Gaussian wave packet and centred at $\overline{y}>0$, with an initial momentum $\overline{p}$, and its mirror image\footnote{The constant $a$ comes from the normalisation of $|\psi_0|^2$ along the positive real axis.}:
\begin{eqnarray}
  \label{initGWP_1D_half}
{  \psi_0(y)=\frac{a}{\sqrt{\sigma}}\left[\ee^{-\frac{(y-\overline{y})^2}{4\sigma^2}+\ic \frac{\overline{p} y}{\hbar}}+
\ee^{-\frac{(y+\overline{y})^2}{4\sigma^2}-\ic \frac{\overline{p} y}{\hbar}}\right]\ ,}\\ 
a=\frac{1}{(2\pi)^{1/4}}\left[1+\ee^{-2(\overline{p}\sigma/\hbar)^2}\left(
1-\re\; \erf\left(-\ic\dfrac{{2}\overline{p}\sigma}{\hbar}\right)\right)\right]^{-1/2}\nonumber\ .
\end{eqnarray}
We introduced the error function defined by:
\begin{equation}
  \erf(Y)=\dfrac{2}{\sqrt{\pi}}\int_0^Y \ee^{-u^2} \ud u\  .
\end{equation}
Note that for a large center position $\overline{y}\gg \sigma$, the contribution of the mirror image is small, and the whole wave function is well approximated by a Gaussian wave packet.
Assume further that $\overline{p}<0$ so that the particle guided by the wave (\ref{initGWP_1D_half}) is sent towards the wall: the centre of the wave packet will first travel towards the wall then be reflected. The quantum time propagator can be found by taking a suitable superposition of free propagators:
\begin{equation}
  \label{prop_wall_1D_N}
  K^{(1D)}_N(y,t|z,0)=\sqrt{\dfrac{m}{2\pi\ic\hbar t}}\left[\ee^{\ic m\frac{(y-z)^2}{2\hbar t}}+ \ee^{\ic m\frac{(y+z)^2}{2\hbar t}}\right], \quad y,z>0 \ .
\end{equation}
Using linearity and {(\ref{GWP_1D})}, the propagated wave packet is:
\begin{multline}
  \label{GWP_wall_1D}
  \psi_N(y,t)=\dfrac{a\;\ee^{\ic\frac{\overline{p}\,\overline{y}}{\hbar}}}{\sqrt{\sigma\left(1+\frac{\ic \hbar t}{2m\sigma^2}\right)}}
\exp\left\{ \dfrac{\ic m}{2 \hbar t}\left[ (y-\overline{y})^2 -\dfrac{(y-\overline{y}-\frac{\overline{p} t}{m})^2}{1+\frac{\ic \hbar t}{2m\sigma^2}}\right] \right\}\\
+\dfrac{a\;\ee^{\ic\frac{\overline{p}\, \overline{y}}{\hbar}}}{\sqrt{\sigma\left(1+\frac{\ic \hbar t}{2m\sigma^2}\right)}}
\exp\left\{ \dfrac{\ic m}{2 \hbar t}\left[ (y+\overline{y})^2 -\dfrac{(y+\overline{y}+\frac{\overline{p} t}{m})^2}{1+\frac{\ic \hbar t}{2m\sigma^2}}\right] \right\}\ .
\end{multline}
It is more convenient to rewrite this expression in the form:
\begin{equation}
  \label{GWP_wallN_1D_v2}
  \psi_N(y,t)=R(t) \ee^{\ic \varphi(y,t)} \left( \ee^{\ic s(y,t)/\hbar}+\ee^{-\ic s(y,t)/\hbar} \right)\ ,
\end{equation}
with the following definitions:
\begin{eqnarray}
  R(t)&=&\dfrac{a\;\ee^{\ic\frac{\overline{p}\,\overline{y}}{\hbar}}}{\sqrt{\sigma\left(1+\frac{\ic \hbar t}{2m\sigma^2}\right)}}\ ,\nonumber\\
\varphi(y,t)&=& \dfrac{m}{2 \hbar t}\left[ (y^2+\overline{y}^2) -\dfrac{y^2+(\overline{y}+\frac{\overline{p} t}{m})^2}{1+\frac{\ic \hbar t}{2m\sigma^2}}\right]\ ,\nonumber\\
s(y,t)&=&\dfrac{\overline{p} y-\ic \dfrac{\hbar t}{2m\sigma^2}\dfrac{m \overline{y} y}{t} }{1+\ic\dfrac{\hbar t}{2m\sigma^2}}\ .\label{S_1D_wall}
\end{eqnarray}
The expression (\ref{GWP_wallN_1D_v2}) enables one to compute the gradient more easily:
\begin{equation}
   \frac{\partial}{\partial y}\psi_N(y,t)=\ic\left[\frac{\partial}{\partial y}\varphi(y,t)\right]\psi_N(y,t)+\frac{\ic}{\hbar}\left[\frac{\partial}{\partial y}s(y,t)\right] R(t) \ee^{\ic \varphi(y,t)} \left( \ee^{\ic s(y,t)/\hbar}-\ee^{-\ic s(y,t)/\hbar} \right) \ .
\end{equation}
Using the definition (\ref{defvpsi}), the Bohmian velocity field is
\begin{equation}
  \label{Bohm_vfield_GWP_wall_1D_N}
  v_N^\psi(y,t)=\dfrac{\left(\dfrac{\hbar t }{2m\sigma^2}\right)^2}{1+\left(\dfrac{\hbar t }{2m\sigma^2}\right)^2}\dfrac{y}{t}+
\re\left\{\dfrac{ \dfrac{\overline{p}}{m} - \ic \dfrac{\hbar t}{2m\sigma^2}\dfrac{\overline{y}}{t}}{1+\ic\dfrac{\hbar t }{2m\sigma^2}}
\dfrac{\ee^{\ic s(y,t)/\hbar}-\ee^{-\ic s(y,t)/\hbar}}{\ee^{\ic s(y,t)/\hbar}+\ee^{-\ic s(y,t)/\hbar}}
\right\}\ .
\end{equation}
This can be compared to the Bohmian velocity field obtained directly from the propagator (\ref{prop_wall_1D_N}):\footnote{Contrarily to what might perhaps be expected, the position moves to the wall rather than to $\overline{y}$ in the limit $t \to 0 $.}
\begin{equation}
  \label{Bohm_vfield_wall_1D}
  v^\psi(y,t)=\dfrac{y}{t}\ .
\end{equation}
This extreme case can also be recovered by taking the limit $\sigma^2$ going to $0$ in (\ref{Bohm_vfield_GWP_wall_1D_N}). Note that, in (\ref{Bohm_vfield_wall_1D}), the Bohmian particle is immediately repelled from the wall (remind that $y>0$). Both expressions (\ref{Bohm_vfield_GWP_wall_1D_N}) and (\ref{Bohm_vfield_wall_1D})  agree at large distance and long time. 

The same method can be used for the case of Dirichlet boundary conditions at the wall $y=0$. The quantum time propagator, similar to (\ref{prop_wall_1D_N}), is now:
\begin{equation}
  \label{prop_wall_1D_D}
  K^{(1D)}_D(y,t|z,0)=\sqrt{\dfrac{m}{2\pi\ic\hbar t}}\left[\ee^{\ic m\frac{(y-z)^2}{2\hbar t}}- \ee^{\ic m\frac{(y+z)^2}{2\hbar t}}\right], \quad y,z>0\ ,
\end{equation}
and the initial Gaussian wave packet, given by the odd version of (\ref{initGWP_1D_half}), evolves as:
\begin{equation}
  \label{GWP_wallD_1D_v2}
  \psi_D(y,t)=R(t) \ee^{\ic \varphi(y,t)} \left( \ee^{\ic s(y,t)/\hbar}-\ee^{-\ic s(y,t)/\hbar} \right)\ ,
\end{equation}
with the same notation as above.
Following the same derivation, the Bohmian velocity field can be written:
\begin{equation}
  \label{Bohm_vfield_GWP_wall_1D_D}
  v_D^\psi(y,t)=\dfrac{\left(\dfrac{\hbar t }{2m\sigma^2}\right)^2}{1+\left(\dfrac{\hbar t }{2m\sigma^2}\right)^2}\dfrac{y}{t}+
\re\left\{\dfrac{ \dfrac{\overline{p}}{m} - \ic \dfrac{\hbar t}{2m\sigma^2}\dfrac{\overline{y}}{t}}{1+\ic\dfrac{\hbar t }{2m\sigma^2}}
\dfrac{\ee^{\ic s(y,t)/\hbar}+\ee^{-\ic s(y,t)/\hbar}}{\ee^{\ic s(y,t)/\hbar}-\ee^{-\ic s(y,t)/\hbar}}
\right\}
\end{equation}
with the same notation as before.

\subsection{Bohmian dynamics in two dimension: One particle in the presence of a wall}

The results obtained in the previous paragraph can be directly used in order to display the trajectory of a Bohmian particle moving in the two$-$dimensional Euclidean plane, in the presence of a wall located at $y=0$ in Cartesian coordinates.
The wave function is \old{initially Gaussian, centred at the point with Cartesian coordinates $(\overline{x},\overline{y})$.}
{a product of a Gaussian wave packet in the $x-$direction, centred at $\overline{x}$ with initial momentum $\overline{p_x}$, and of the symmetrised Gaussian state as in (\ref{initGWP_1D_half}) in the $y-$direction.}
It has the same width $\sigma$ in both $x-$ and $y-$directions.\old{, and an initial momentum with Cartesian coordinates $(\overline{p_x},\overline{p_y})$}. Using {(\ref{Bohm_vfield_GWP_1D})} for the dynamics along the $x$ direction, and (\ref{Bohm_vfield_GWP_wall_1D_N}) or (\ref{Bohm_vfield_GWP_wall_1D_D}) along the $y-$direction, the Bohmian velocity field has the Cartesian coordinates $(v_x^\psi,v_y^\psi)$ with:
\begin{eqnarray}
  v_x^\psi&=&\dfrac{\dfrac{\overline{p_{x}}}{m}+\left(\dfrac{\hbar t}{2m\sigma^2}\right)^2 \dfrac{x-\overline{x}}{t}}{1+\left(\dfrac{\hbar t}{2m\sigma^2}\right)^2}\ ,\label{vx_wall}\\
  v_y^\psi&=&\dfrac{\left(\dfrac{\hbar t }{2m\sigma^2}\right)^2}{1+\left(\dfrac{\hbar t }{2m\sigma^2}\right)^2}\dfrac{y}{t}+
\re\left\{\dfrac{ \dfrac{\overline{p_y}}{m} - \ic \dfrac{\hbar t}{2m\sigma^2}\dfrac{\overline{y}}{t}}{1+\ic\dfrac{\hbar t }{2m\sigma^2}}
\dfrac{\ee^{\ic s(y,t)/\hbar}-\varepsilon \ee^{-\ic s(y,t)/\hbar}}{\ee^{\ic s(y,t)/\hbar}+\varepsilon \ee^{-\ic s(y,t)/\hbar}}\ ,
\right\} \label{vy_wall}
\end{eqnarray}
where $\varepsilon=+1$ (resp. $-1$) for Neumann (resp. Dirichlet) boundary condition on the wall {and} $s(y,t)$ is given by (\ref{S_1D_wall}).

\begin{figure}[!h]
  \begin{minipage}[l]{0.49\linewidth}
    \begin{center}
      \includegraphics[width=\textwidth]{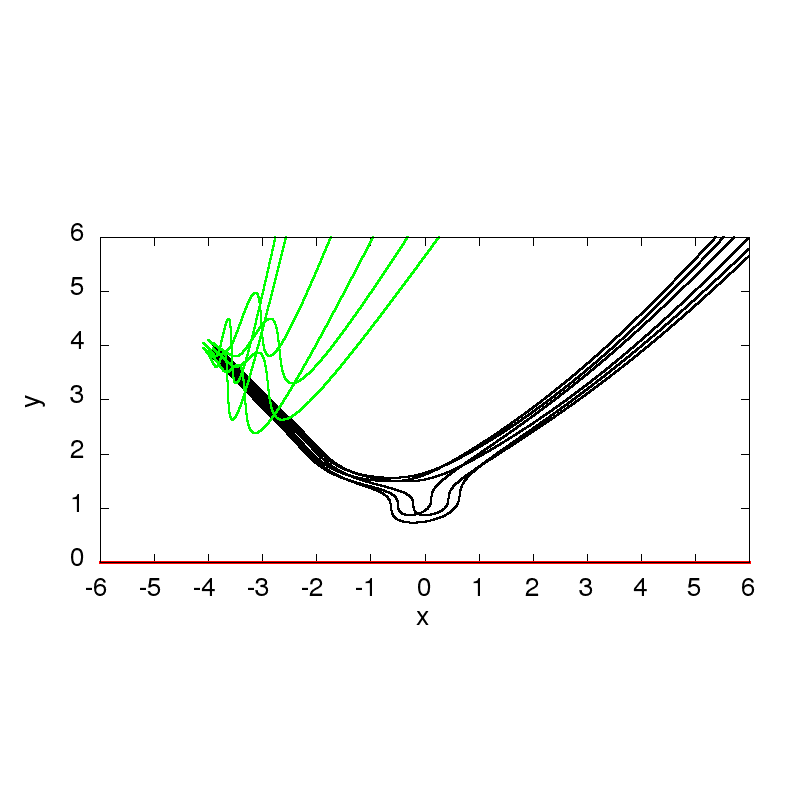}
    \end{center}
  \end{minipage}
  \begin{minipage}[r]{0.49\linewidth}
    \begin{center}
      \includegraphics[width=\textwidth]{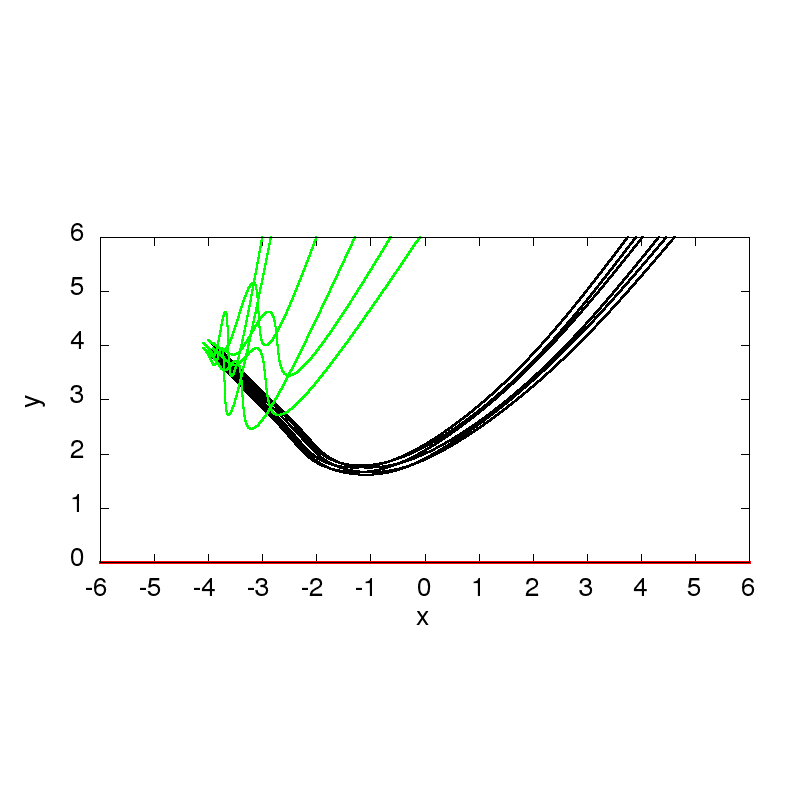}
    \end{center}
  \end{minipage}
  \caption{
Bohmian trajectories of a 2D particle hitting a wall located at $y=0$ {with Neumann boundary conditions}. The initial state is a Gaussian wave packet centred $(\overline{x},\overline{y})$, with width $\sigma$ and an initial momentum $(\overline{p_x},\overline{p_y})$. The initial positions of the Bohmian trajectories are along a circle centred at $(\overline{x},\overline{y})$ with radius $\rho$ at time $t_{\rm init}$. Here: $(\overline{x},\overline{y})=(-4,4)$, $\sigma=1,\ \rho=0.1$. The magnitude of the initial momentum is $|{\bf \overline{p}}|$ and its direction in polar coordinates is $\theta_0=\pi/4$ such that $\overline{p_x}=\overline{p_y}=|{\bf \overline{p}}|/\sqrt{2}$. Black: $|{\bf \overline{p}}|=\frac{\hbar}{2\sigma}\frac{\overline{y}}{\sigma}$. Green: $|{\bf \overline{p}}|=\frac{\hbar}{2\sigma}\frac{\sigma}{\overline{y}}$.
Left: $t_{\rm init}=0.01$. Right: $t_{\rm init}=0.1$. 
} 
\label{GaussWP_Neumann_wall}
\end{figure}

\begin{figure}[!h]
  \begin{minipage}[l]{0.49\linewidth}
    \begin{center}
      \includegraphics[width=\textwidth]{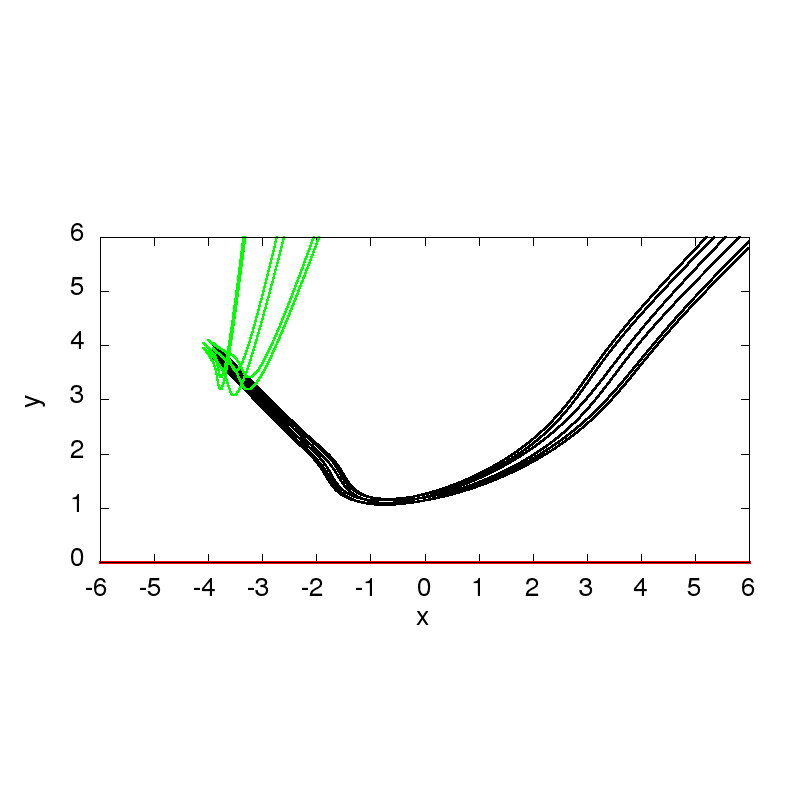}
    \end{center}
  \end{minipage}
  \begin{minipage}[r]{0.49\linewidth}
    \begin{center}
      \includegraphics[width=\textwidth]{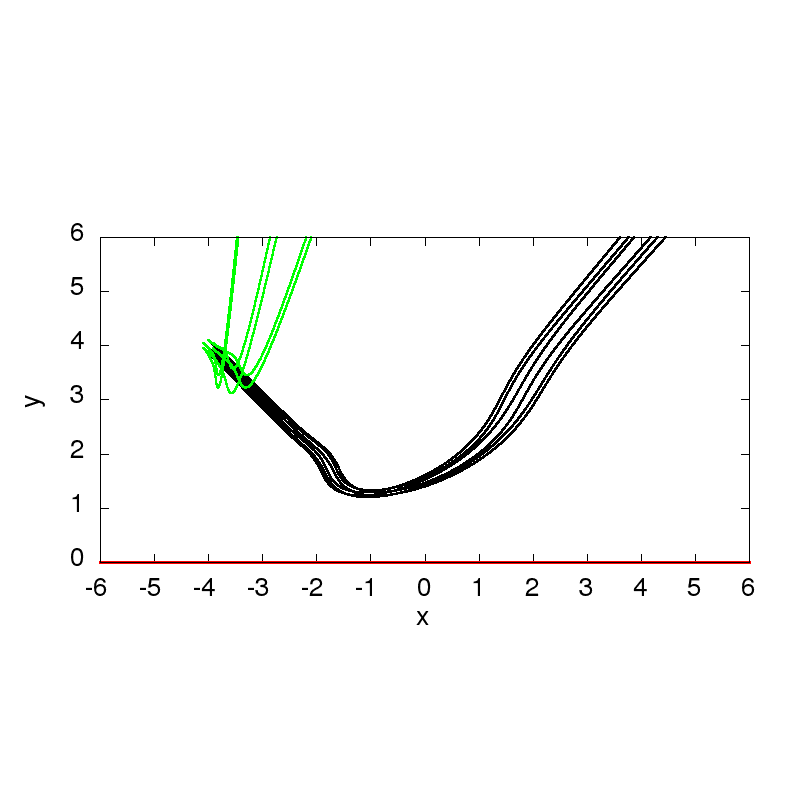}
    \end{center}
  \end{minipage}
  \caption{Same as Fig.~\ref{GaussWP_Neumann_wall} with Dirichlet boundary conditions.} 
\label{GaussWP_Dirichlet_wall}
\end{figure}

We computed the trajectories for Neumann boundary conditions, see Fig.~\ref{GaussWP_Neumann_wall}, and {for Dirichlet boundary conditions at the wall, see Fig.~\ref{GaussWP_Dirichlet_wall}}.
The first observation is that, as expected, the trajectory of a Bohmian particle guided by a wave packet can be close to the classical trajectory of the corresponding Hamiltonian classical system. Such a similarity is reached by taking an initial wave packet with a small width ($\sigma\ll |\overline{y}|$), and an initial momentum magnitude $|{\bf \overline{p}}|$ such that the real and the imaginary parts of the first factor of the numerator in the last term of (\ref{vy_wall}) are of the same magnitude, i.e. $|\overline{p_y}|=\frac{\hbar |\overline{y}|}{2\sigma^2}\gg \frac{\hbar}{\sigma}$. This can be equivalently described in words as the regime where the momentum associated with the motion of the centre of the initial wave packet is much larger that the intrinsic momentum describing the spreading of the initial wave packet. 
\old{In the case of Neumann boundary conditions on the wall, the Bohmian particle goes closer to the wall than in the case of Dirichlet boundary conditions.}

For sake of comparison, we also displayed the Bohmian trajectories for the same initial and times when the ``spreading'' momentum is much larger than the initial classical momentum in Fig.~\ref{GaussWP_Neumann_wall} and Fig.~\ref{GaussWP_Dirichlet_wall} (green curves). The similarity with classical dynamics is less obvious then.

Trajectories may also display different behaviour for different initial positions. Those differences are less important for Dirichlet than for Neumann boundary conditions at the wall. This can be seen by comparing the figures on the left with those on the right. Rather than varying the initial positions, the initial time at which a fixed position is considered, was varied.

Finally, another difference with the classical dynamics is that classical particles will actually hit the wall, which causes a discontinuity in their velocity, while the Bohmian trajectories do not hit the wall and have smooth trajectories.

\section{Bohmian particle in the presence of a half-line barrier}\label{half-line}

In the context of classical optics, the problem of a half-line barrier is well-studied. The solution is discussed in detail in \cite{sommerfeld96}. In the context of quantum mechanics, the solution by means of the propagator for both Dirichlet and Neumann boundary conditions was provided by Schulman \cite{schulman82}. It was revisited within the path integral approach in \cite{dewitt-morette86}. The half-line barrier scattering problem is somewhat exceptional since it is one of the cases for which the propagator is a simple expression in terms of special functions. Already for the scattering by a single slit, the propagator is much more complicated as it is given by a Fourier integral of a series of special (Mathieu) functions{, see e.g.~\cite{MorseFeshbach2}}.

Similarly to the previous section three types of wave functions can be considered.
One may indeed consider the time propagator, for which the initial wave function is a Dirac delta distribution.
In order to improve the readability, the corresponding trajectories are displayed in \ref{Traj_quantum_prop}.
Below we display a more realistic example of an initial Gaussian wave packet. In the case of zero initial momentum of the Gaussian packet, the trajectories agree very well qualitatively with those corresponding to the propagator. We also consider an initial Gaussian wave packet with non-zero initial momentum. Moreover we will consider a stationary state corresponding to an incoming plane wave. Such a state is sometimes called a scattering state in scattering theory, see e.g.~\cite{landau58}. This last wave function is especially suitable for direct comparison with the dynamics of a walker.

\subsection{Quantum time propagator}
{We consider the half-line barrier ${\cal O}$ along the negative $x$-axis, i.e., 
\begin{equation}
  \label{obst}
  {\cal O}: y=0, x\le 0\ .
\end{equation}
The propagator for the half-line barrier is \cite{schulman82}:}
\begin{equation}
  \label{prop_halflineN}
K_{N,D}({\bf x},t|{\bf x}_0,0)=\dfrac{m}{2\pi\ic \hbar t} \ee^{\ic\frac{ m (r+r_0)^2}{2\hbar t}}\left[ F\left(u_1\right) \pm
F\left(u_2\right)\right]\ ,
\end{equation}
where $(r,\theta)$ and $(r_0,\theta_0)$ respectively {denote} the polar coordinates of ${\bf x}$ and ${\bf x}_0$, with $-\pi\le \theta,\theta_0 < \pi$, 
\begin{equation}
  u_1 =\sqrt{\dfrac{2m r r_0}{\hbar t}} \cos\left(\dfrac{\theta-\theta_0}{2}\right) \ , \qquad u_2 = -\sqrt{\dfrac{2m r r_0}{\hbar t}} \cos\left(\dfrac{\theta+\theta_0}{2}\right) \ ,
\label{u1u2}
\end{equation}
and
\begin{equation}
  \label{defF}
  F(u)=\frac{1}{\sqrt{\pi}} \ee^{-\ic u^2 - \ic \pi/4}\int_{-\infty}^u \ee^{\ic v^2}\ud v\ .
\end{equation}
Neumann boundary conditions, indicated by the index $N$, give the relative plus sign, whereas the Dirichlet boundary conditions (index $D$) give the relative minus sign.

\subsection{Initial Gaussian wave packet}

We consider the following initial Gaussian wave packet:
\begin{equation}
  \label{initGWP}
  \psi_0({\bf x})=\frac{1}{2\pi\sigma^2}\ee^{-\frac{({\bf x}-{\bf \overline{x}})^2}{4\sigma^2}+\ic \frac{{\bf \overline{p}} {\bf x}}{\hbar}}\ .
\end{equation}
We did not find a way to compute the expression the wave function and its gradient, which are respectively given by (\ref{GWP_2D}) and 
\begin{equation}
\nabla \psi({\bf x}, t) = \int \left[ \nabla K_{N,D}({\bf x},t|{\bf x}_0,0)\right] \psi_0({\bf x}_0) \ud \mathbf{x}_0\ ,\label{dpsiint}
\end{equation}
where the gradient of the propagator is given in \ref{dK}, analytically (neither for Neumann nor Dirichlet boundary conditions). Consequently we did not find an analytic expression for the Bohmian velocity. Therefore, to evaluate the wave function and its gradient, a numerical integration by means of the Gaussian-quadrature algorithm was employed. 
In this numerical integration, the spatially localised feature of the initial wave function was exploited for the efficiency of the numerical computation. As the initial wave function is given by a Gaussian wave packet, the integration could be performed in the finite region $[\overline{x}-3\sigma, \overline{x}+3\sigma]\times[\overline{y}-3\sigma, \overline{y}+3\sigma]$ instead of the whole space\footnote{That is the reason why we do not worry about imposing the actual boundary conditions in Eq.~(\ref{initGWP}).}. This guarantees faster computation with a better accuracy. 
The initial Bohmian positions were taken circularly around the centre of the initial Gaussian wave packet with radius {$\rho$ at some time $t_{\rm init}$}. For the numerical integration of the Bohmian velocity, the 4th-order Runge-Kutta was implemented.

\begin{figure}[!ht]
  \begin{minipage}[l]{0.49\linewidth}
    \begin{center}
      \includegraphics[width=\textwidth]{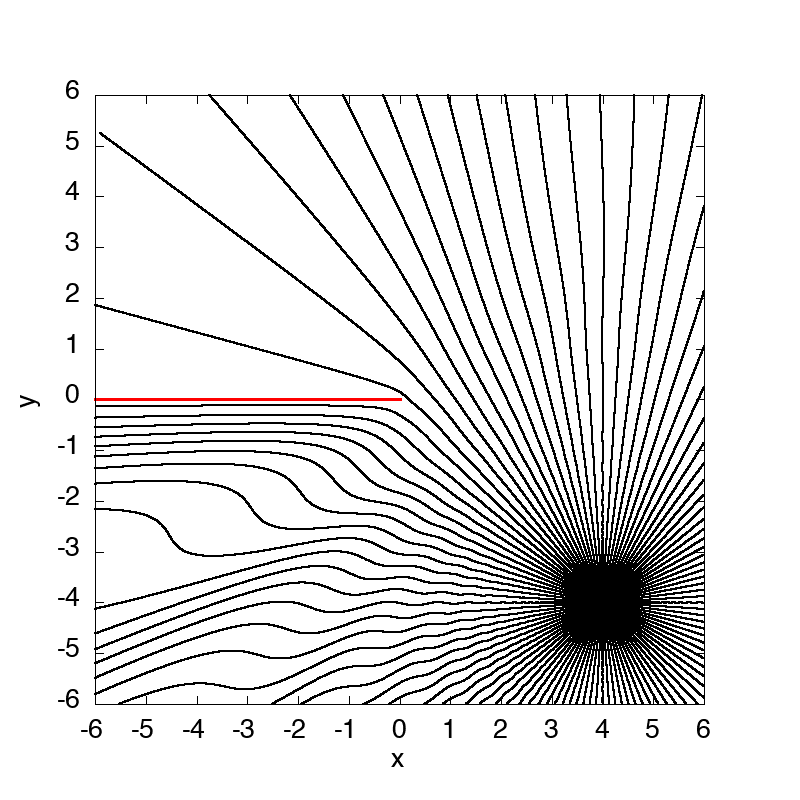}
    \end{center}
  \end{minipage}
  \begin{minipage}[r]{0.49\linewidth}
    \begin{center}
      \includegraphics[width=\textwidth]{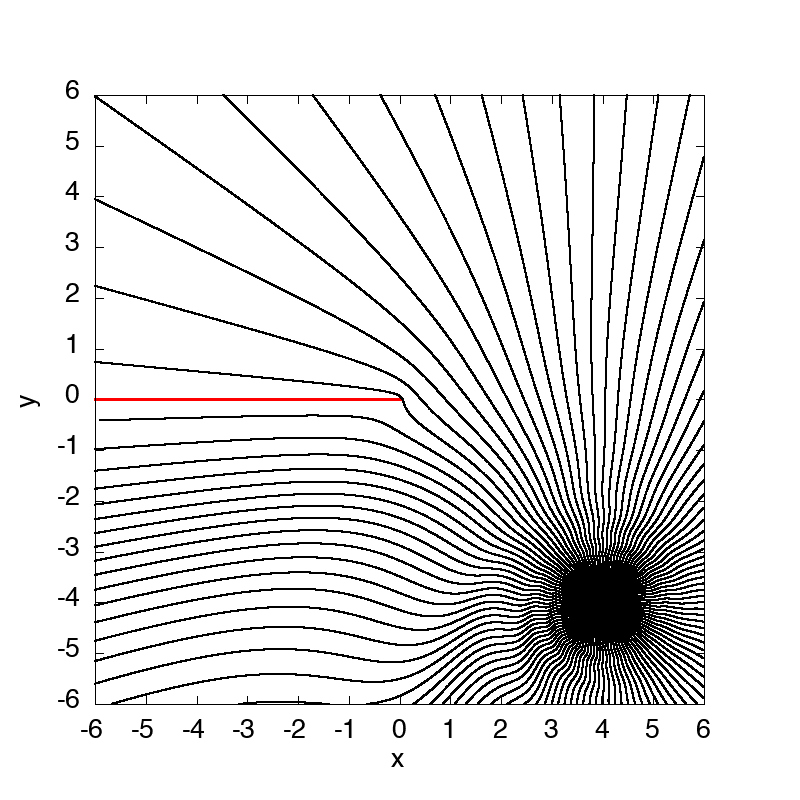}
    \end{center}
  \end{minipage}
  \caption{Bohmian trajectories in the presence of a half-line obstacle with Neumann boundary conditions. The initial wave function is the Gaussian (\ref{initGWP}) centred at $(\overline{x},\overline{y})=(4,-4)$, with $\sigma=0.1$ and ${\bf \overline{p}}={\bf 0}$. The initial positions are along a circle centred at $(\overline{x},\overline{y})$ with radius $\rho=0.02$ at time $t_{\rm init}$. Left: {$t_{\rm init}=0.01$. Right: $t_{\rm init}=0.05$.}}
\label{GWP_N_x4}
\end{figure}

\begin{figure}[!ht]
  \begin{minipage}[l]{0.49\linewidth}
    \begin{center}
      \includegraphics[width=\textwidth]{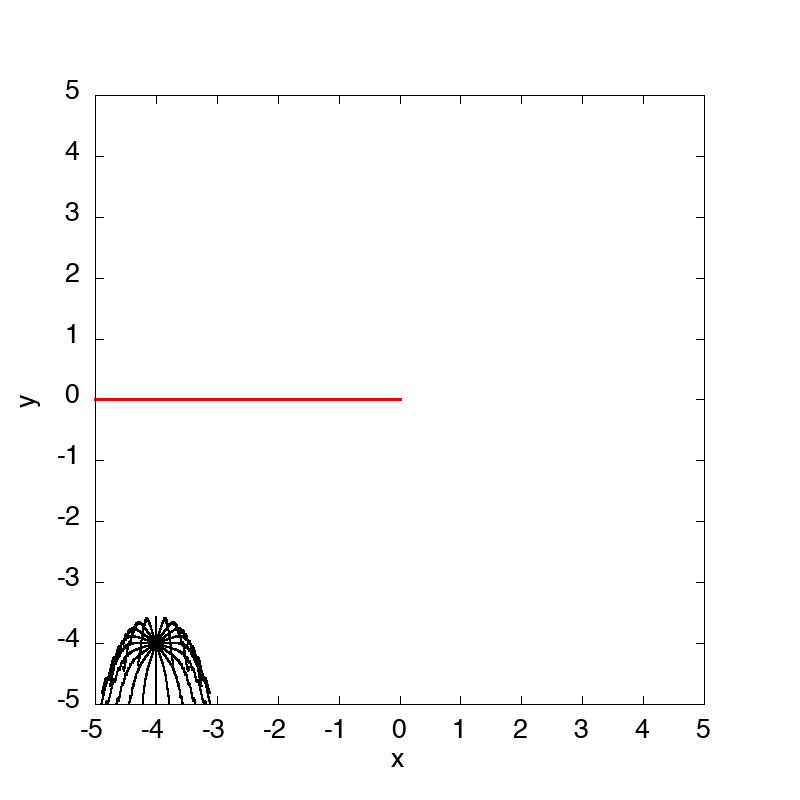}
    \end{center}
  \end{minipage}
  \begin{minipage}[r]{0.49\linewidth}
    \begin{center}
      \includegraphics[width=\textwidth]{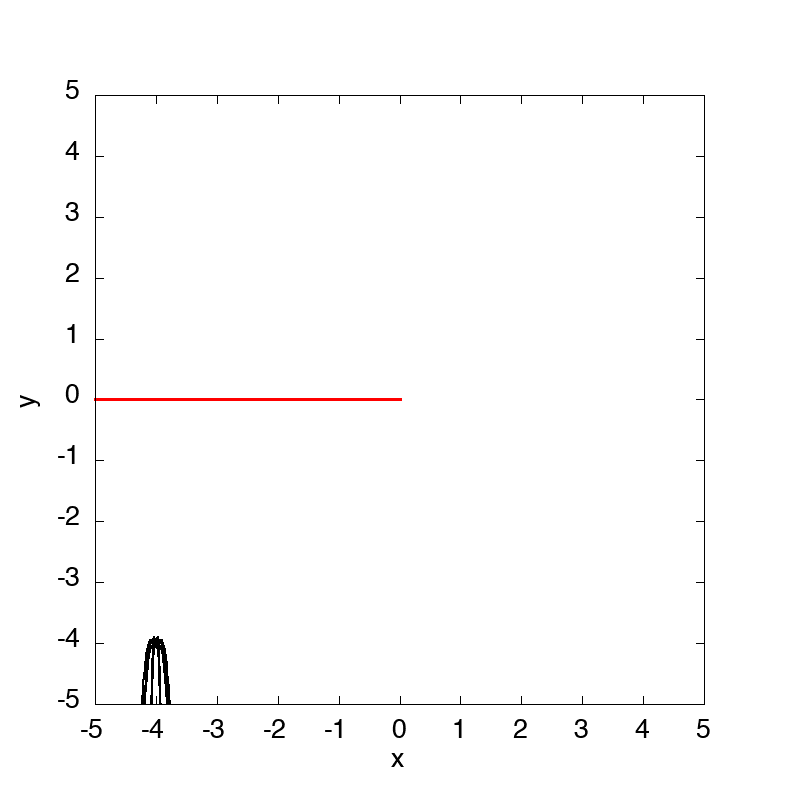}
    \end{center}
  \end{minipage}
  \caption{Bohmian trajectories in the presence of a half-line obstacle with Neumann boundary conditions. {The initial wave function is the Gaussian (\ref{initGWP}) centred at $(\overline{x},\overline{y})=(-4,-4)$, with $\sigma=0.1$ and ${\bf \overline{p}}={\bf 0}$. The initial positions are along a circle centred at $(\overline{x},\overline{y})$ with radius $\rho=0.02$ at time $t_{\rm init}$.} Left: {$t_{\rm init}=0.01$. Right: $t_{\rm init}=0.05$.}}
\label{GWP_N_x-4}
\end{figure}

\begin{figure}[!ht]
  \begin{minipage}[l]{0.49\linewidth}
    \begin{center}
      \includegraphics[width=\textwidth]{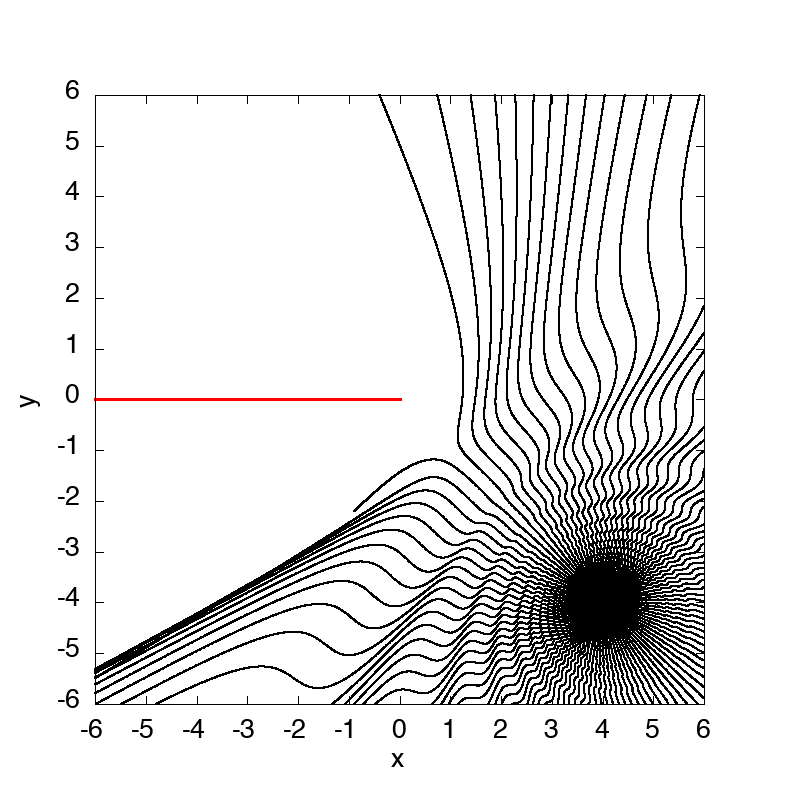}
    \end{center}
  \end{minipage}
  \begin{minipage}[r]{0.49\linewidth}
    \begin{center}
      \includegraphics[width=\textwidth]{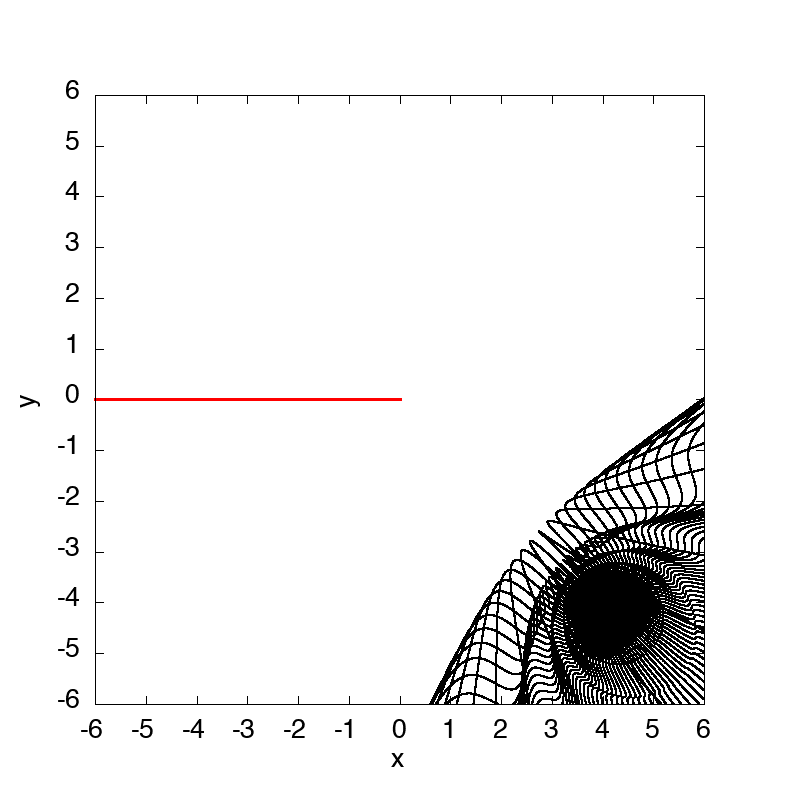}
    \end{center}
  \end{minipage}
  \caption{
Bohmian trajectories in the presence of a half-line obstacle with Dirichlet boundary conditions. The initial wave function is the Gaussian (\ref{initGWP}) centred at $(\overline{x},\overline{y})=(4,-4)$, with $\sigma=0.1$ and ${\bf \overline{p}}={\bf 0}$. The initial positions are along a circle centred at $(\overline{x},\overline{y})$ with radius $\rho=0.02$ at time $t_{\rm init}$. Left: 
{$t_{\rm init}=0.01$. Right: $t_{\rm init}=0.05$.}
}
\label{GWP_D_x4}
\end{figure}

\begin{figure}[!ht]
  \begin{minipage}[l]{0.49\linewidth}
    \begin{center}
      \includegraphics[width=\textwidth]{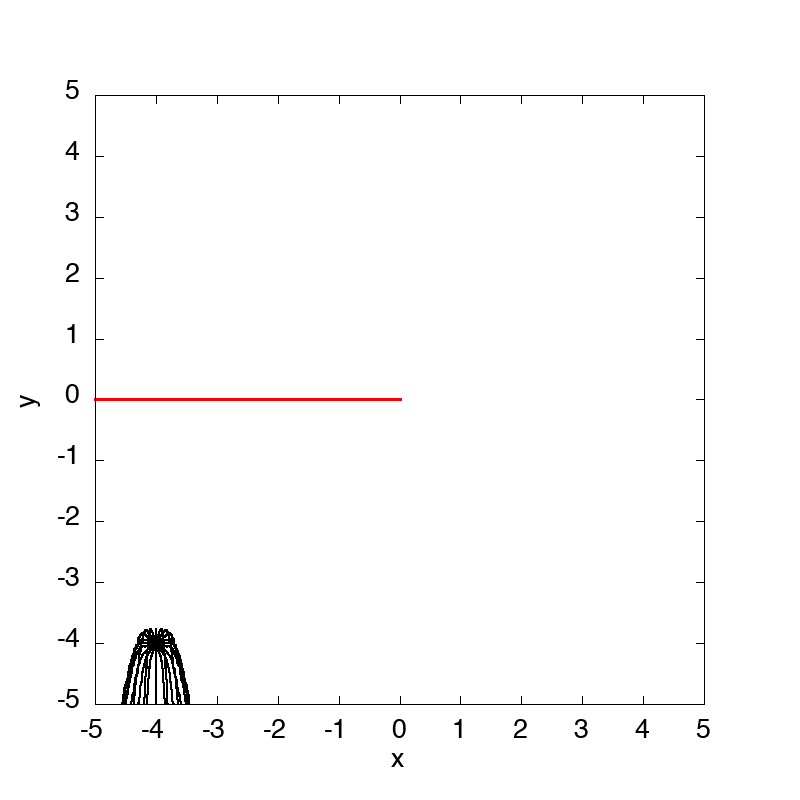}
    \end{center}
  \end{minipage}
  \begin{minipage}[r]{0.49\linewidth}
    \begin{center}
      \includegraphics[width=\textwidth]{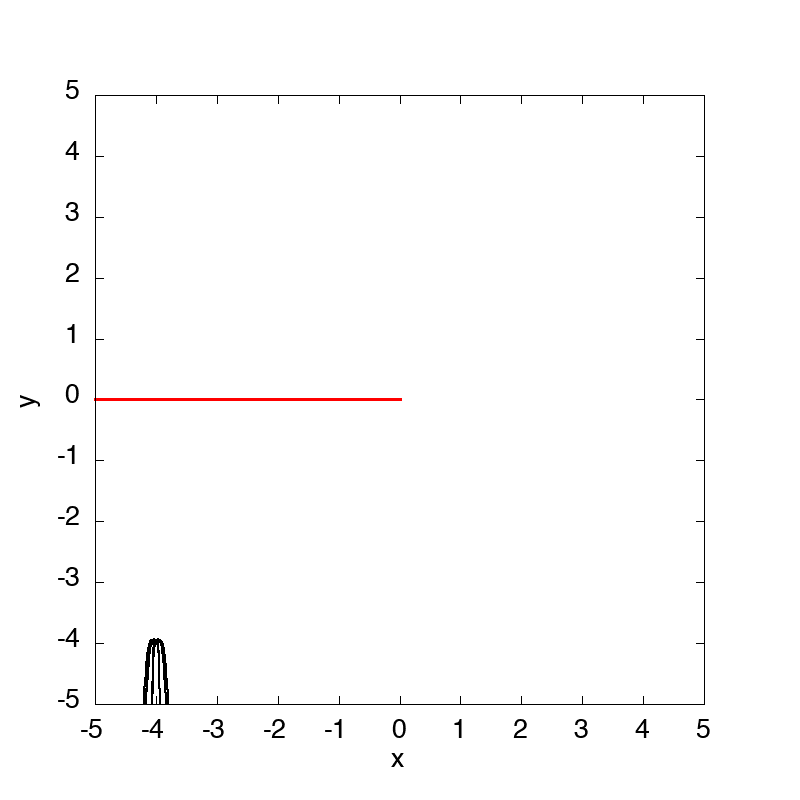}
    \end{center}
  \end{minipage}
  \caption{
Bohmian trajectories in the presence of a half-line obstacle with Dirichlet boundary conditions. The initial wave function is the Gaussian (\ref{initGWP}) centred at $(\overline{x},\overline{y})=(-4,-4)$, with $\sigma=0.1$ and ${\bf \overline{p}}={\bf 0}$. The initial positions are along a circle centred at $(\overline{x},\overline{y})$ with radius $\rho=0.01$ at time $t_{\rm init}$. Left: 
{$t_{\rm init}=0.001$. Right: $t_{\rm init}=0.005$.}
}
\label{GWP_D_x-4}
\end{figure}

First, in Figs.~\ref{GWP_N_x4}, \ref{GWP_N_x-4}, \ref{GWP_D_x4} and \ref{GWP_D_x-4}, the Bohmian trajectories are computed {for a wave function with zero initial momentum}, i.e.\ ${\bf \overline{p}}={\bf 0}$. 
One can see strong similarities with the trajectories computed from the propagator, see the figures in \ref{Traj_quantum_prop}. This may be viewed as a benchmark: in the case of an initially highly localised wave packet ($\sigma\ll 1$, and $\sigma\ll |{\bf \overline{x}}|$) it is reasonable to compute the Bohmian trajectories from the quantum propagator directly (unlike the case of non-zero initial momentum). 

In the cases \old{where }the initial point sits in a region where at large distance, the wave function is a superposition of the incident wave and the reflected wave (i.e.\ the initial point is the third quadrant in our choice of coordinates), see Figs.~\ref{GWP_N_x-4}, \ref{GWP_D_x-4}, the Bohmian trajectories are similar to those of a particle in the presence of a wall and the diffractive effects are less important.

Comparing Fig.~\ref{GWP_N_x4}~right and Fig.~\ref{GWP_D_x4}~right, there is a very important difference between Neumann and Dirichlet boundary conditions at the barrier. What is seen in Fig.~\ref{GWP_N_x4}~right for Neumann boundary conditions is that the particle is not repelled by the barrier but may flow next to it. In Fig.~{\ref{GWP_D_x4}}~right which concerns Dirichlet boundary conditions, the particle is repelled strongly at \emph{moderate} distances from the tip. Similar behaviour was observed in the case of the particle in the half-plane. The reason is that in the case of Dirichlet boundary conditions the probability density $|\psi|^2$ is zero at the barrier. Hence due to the low density near the barrier, the particle tends to avoid that region. 

Note that the {particles move radially} at \emph{large enough} distance, in agreement with our analytical asymptotic formulas derived in \ref{farfield} for the propagator, which are valid at large distance at any \emph{fixed} time. 

\begin{figure}[!ht]
    \begin{center}
      \includegraphics[width=0.5\textwidth]{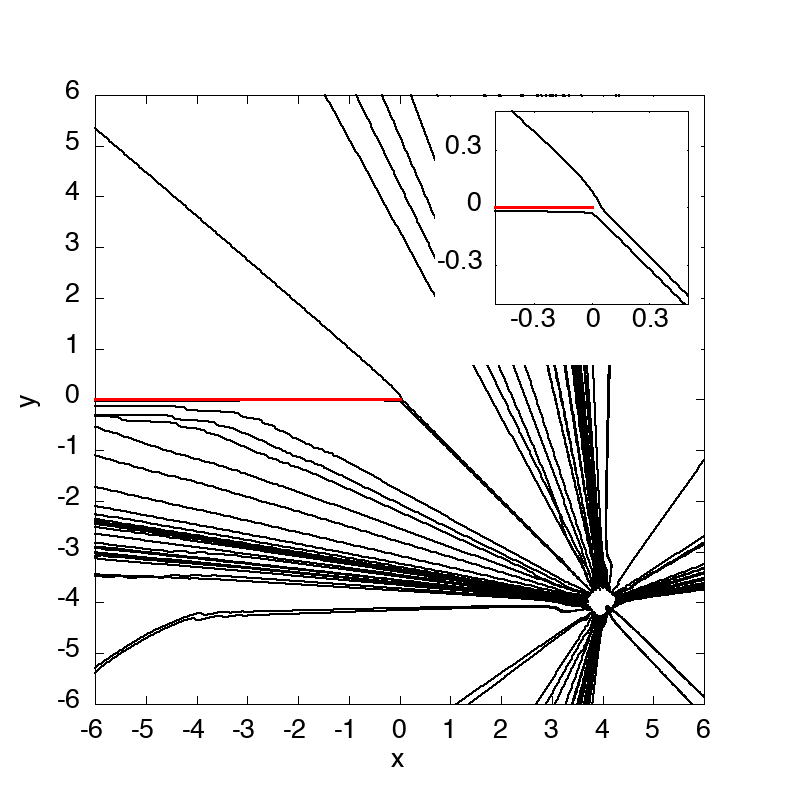}
    \end{center}
    \caption{ 
Bohmian trajectories in the presence of a half-line obstacle with Dirichlet boundary conditions. The initial wave function is the Gaussian (\ref{initGWP}) centred at $(\overline{x},\overline{y})=(4,-4)$, with $\sigma=0.5$ and 
$|{\bf \overline{p}}|=\frac{\hbar |\overline{y}|}{2\sigma^2}$. The initial positions are along a circle centred at $(\overline{x},\overline{y})$ with radius $\rho=0.25$ at time $t_{\rm init}=0.001$.
Inset: As Dirichlet boundary conditions are assigned in this computation, the repulsion of a Bohmian trajectory from the tip is observed in a small scale.}
\label{GWP_initmom}
\end{figure}

Finally, we also computed the Bohmian trajectories for a Gaussian wave packet with a non-zero initial momentum, see Fig.~\ref{GWP_initmom}. It can be seen that as in the case of the half-plane barrier, the Bohmian trajectories are closer to the classical trajectories given by Hamiltonian dynamics.

\subsection{Incoming plane wave. Comparison with walking droplets. }

Recently, a fluid dynamical system has been suggested to have similar characteristic features as Bohmian quantum mechanics \cite{couder06,bush15}. In this system, a droplet is bouncing on the surface of a vibrating fluid bath. Due to the vibration of the bath, the bouncing dynamics can be kept for a long time if the system (vibration frequency and amplitude, droplet size,\dots) is carefully prepared. At each bounce, the droplet emits a standing surface wave due to Faraday instability, which acts as a propelling wave for the droplet at later time. The behaviour of droplet and the surface wave seem similar to that of the particle and wave function in Bohmian mechanics. As such this fluid dynamical system was investigated as a possible classical analogue of quantum mechanics. One fascinating aspect is that one can observe the motion of the droplet and the surface wave without disturbing them, whereas in Bohmian mechanics an observation of the trajectories causes strong disturbance.

In the fluid dynamical system, the depth of the vibrating bath can vary. If it is small enough, the dispersion relation is changed so that no surface Faraday wave can be emitted and the droplet is effectively expelled from this region. This is how a barrier has been engineered in the experiment. In the experiment detailed in \cite{harris15}, a walking droplet is sent towards a half-line barrier with {an initial velocity in} the $y-$direction\footnote{The position of the obstacle and the choice of coordinates are the same as in (\ref{obst}).}. 

\begin{figure}[!ht]
    \begin{center}
      \includegraphics[width=0.5\textwidth]{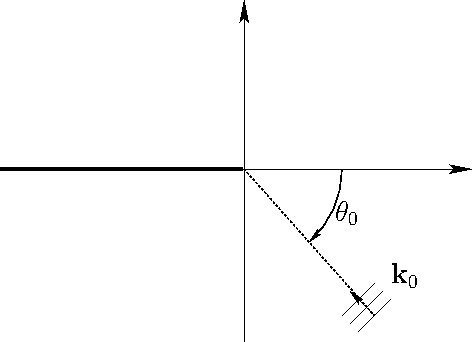}
    \end{center}
\caption{Choice of coordinates for an incoming plane wave.}\label{def_t0}
\end{figure}

\begin{figure}[!ht]
  \begin{minipage}[l]{0.49\linewidth}
    \begin{center}
      \includegraphics[width=\textwidth]{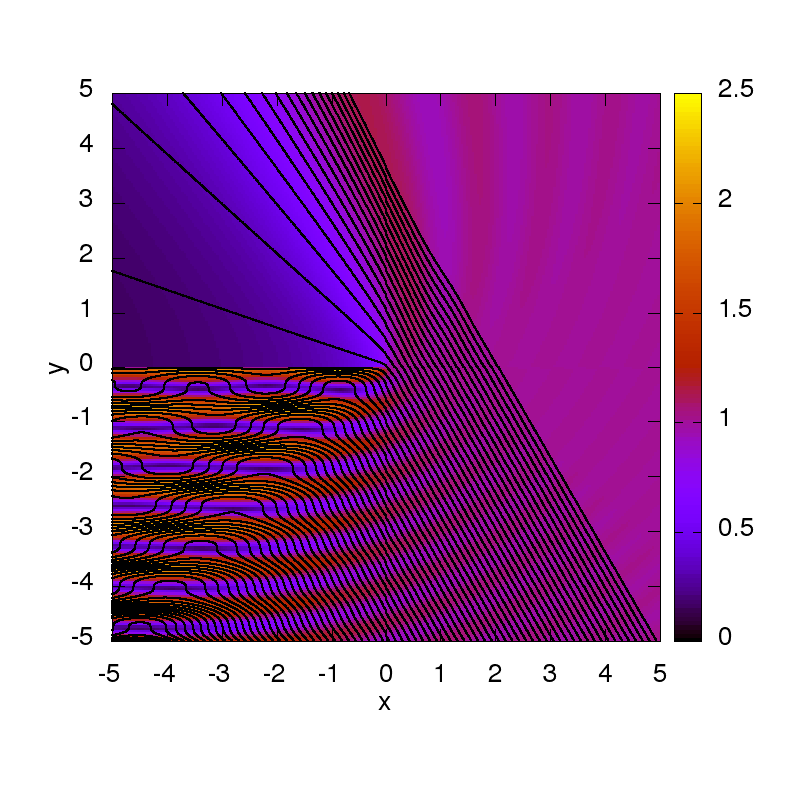}
    \end{center}
  \end{minipage}
  \begin{minipage}[r]{0.49\linewidth}
    \begin{center}
      \includegraphics[width=\textwidth]{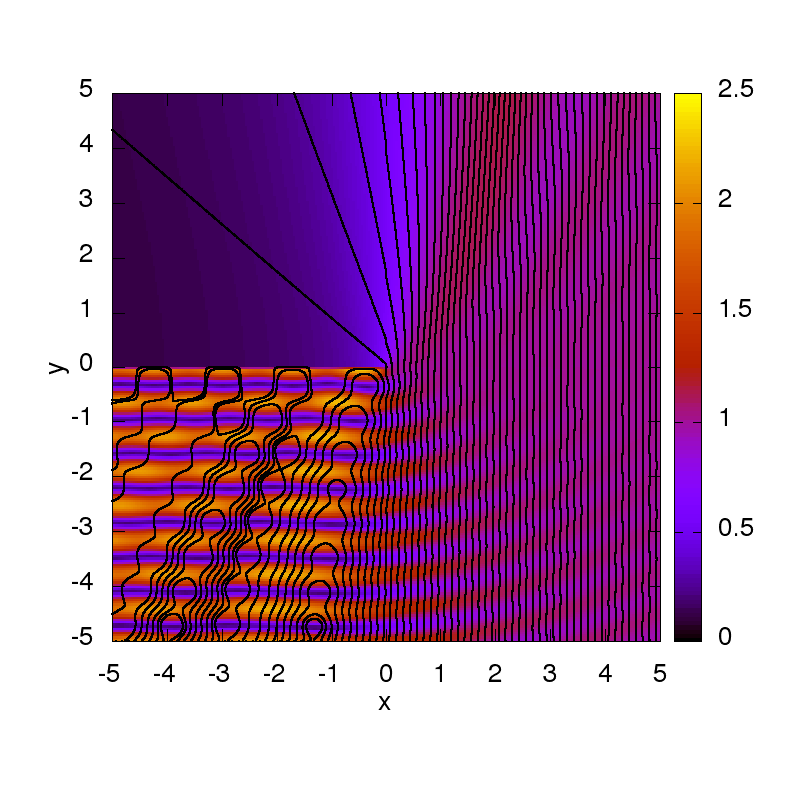}
    \end{center}
  \end{minipage}
  \caption{Bohmian trajectories in the presence of a half-line obstacle with Neumann boundary conditions. The velocity field is computed {for} an incident plane wave (\ref{scat_state}) with $k_0=5$. The background colour map stands for the modulus of the scattering state (\ref{scat_state}). Left: $\theta_0=\pi/3$. Right: $\theta_0=\pi/2$.}
\label{Bohm_traj_pw}
\end{figure}

The analogous situation for a Bohmian particle is to consider an initial plane wave, so that the initial momentum is fixed to the $y-$direction. It amounts to solve the scattering problem for an incoming plane wave with momentum $\hbar {\bf k}_0$. Therefore one needs to solve the Schr\"odinger equation (\ref{schro}) for a fixed energy $E=\hbar^2 {\bf k}^2_0/2m$, which reduces to solve Helmholtz equation:
\begin{equation}
  \label{helm}
  (\Delta+k_0^2)\psi=0\ ,
\end{equation}
This equation needs to be supplemented with boundary conditions on the obstacle and at infinity. At infinity the wave is the sum of the incident plane wave and a scattered wave. The initial momentum ${\bf k}_0$ is assumed to have the Cartesian coordinates {$(k_x,k_y)=-(k_0\cos\theta_0,k_0\sin\theta_0)$} with $\theta_0$ defined in Fig.~\ref{def_t0}. 
Along the obstacle we will assume that the surface wave obeys Neumann boundary conditions as recently proposed in \cite{dubertrand16}.
This is exactly the optical problem, which was solved originally by Sommerfeld \cite{sommerfeld96}. So here we only give the result (see also \cite{sommerfeld64} for a more detailed derivation):
\begin{equation}
   \psi({\bf x})=\ee^{\ic k_0 r}\left[F\left(a_1\right) + F\left(a_2\right)\right]\ ,\label{scat_state}
\end{equation}
with 
\begin{equation}
  a_1 =\sqrt{2k_0 r}\cos\left(\frac{\theta-\theta_0}{2}\right)\ , \qquad  a_2 = -\sqrt{2k_0 r}\cos\left(\frac{\theta+\theta_0}{2}\right)  \ ,
\end{equation}
and $F$ as defined in (\ref{defF}). A simple computation shows that 
\begin{equation*}
  {\boldsymbol \nabla} \psi = \ee^{\ic k_0 r} \Bigg\{ \ic \left[F(a_1) {\bf k_0}+ F(a_2) {\bf k'_0}\right] + \ee^{-\ic\pi/4} \sqrt{\frac{2k_0}{\pi r}} \sin\left(\frac{\theta_0}{2}\right) \left[ -\sin\left(\frac{\theta}{2}\right){\bf e}_x + \cos\left(\frac{\theta}{2}\right){\bf e}_y \right] \Bigg\} \ ,
\end{equation*}
where ${\bf k'_0}=(k_x,-k_y)$. The Bohmian velocity field is then, using (\ref{defvpsi}):
\begin{multline}
{\bf v}^\psi = \frac{\hbar k_x }{m}  {\bf e}_x + \frac{\hbar  k_y}{m} {\bf e}_y \frac{|F(a_1)|^2 -|F(a_2)|^2 }{|F(a_1)+F(a_2)|^2} \\
+ \frac{\hbar}{m}\sqrt{\frac{2k_0}{\pi r}} {\textrm{Im}}\left( \frac{\ee^{-\ic\pi/4}}{F(a_1)+F(a_2)} \right)\sin\left(\frac{\theta_0}{2}\right) \left[ -\sin\left(\frac{\theta}{2}\right){\bf e}_x + \cos\left(\frac{\theta}{2}\right){\bf e}_y \right]  \ .
\end{multline}
Note that the wave function (\ref{scat_state}) is time-independent, {hence so is the Bohmian velocity field}. As such, the initial time assigned to an initial {position} is not relevant. Some Bohmian trajectories corresponding to this wave function are displayed in Fig.~\ref{Bohm_traj_pw}. 
First it is worth observing that the Bohmian particle spends more time in the region where the probability density is higher. It is also avoiding the zeros by circulating around them, see more precisely a \emph{closed} trajectory around {the point $(-1.5,-4.7)$} in Fig.~\ref{Bohm_traj_pw}~right.
Second, 
in the droplet case, see Figs.~5-10\ p.142 in [22], the trajectories passing near the tip seem to have a fixed outgoing direction unlike in the Bohmian case. This is another reason to believe that while both systems certainly have pilot-wave features, they differ significantly in the details of the obtained pattern.

{From the quantum perspective, one may also consider the case of an incoming plane wave in the case of Dirichlet boundary conditions on the half-line barrier. The analysis is similar and the results are presented in \ref{pw_D}.}

\section{Conclusion}\label{conclusion}
We have studied the Bohmian trajectories in the case of the half-line barrier with Neumann and Dirichlet boundary conditions. This is one of the simplest examples of a diffractive scattering problem (simpler than e.g.\ the single or double slit set-up). Making use of the quantum time propagator (which is known exactly in this case), we numerically computed the Bohmian trajectories.

We considered wave functions that are initially Gaussian and stationary wave functions, with different momenta. In both cases, the trajectories are repelled more strongly by the barrier in the case of Dirichlet boundary conditions compared to Neumann boundary conditions. Unlike in the classical case, the trajectories do not hit the barrier but rather curve around it.

A comparison was made with the bouncing droplet system in the case of the stationary state with incoming momentum orthogonal to the barrier with Neumann boundary conditions. A clear difference was found in the scattering pattern.  

In future work, the single and double slit setup could be considered using approximate propagators, like considered in \cite{beau12}. Moreover the computation of Bohmian trajectories via the propagator formalism might be helpful for more practical problems, as recently listed in \cite{benseny14}.

\ack

R.D.\ acknowledges fruitful discussion with A.\ Goussev at an early stage of the project. R.D.\ and J.-B.S.\ acknowledge financial support from the `Actions de Recherches Concert\'ees (ARC)' of the Belgium Wallonia Brussels Federation under contract No.~12-17/02. Computational resources have been provided by the Consortium des \'Equipements de Calcul Intensif (C\'ECI), funded by the Fonds de la Recherche Scientifique de Belgique (F.R.S.-FNRS) under Grant No. 2.5020.11. W.S.\ acknowledges financial support from the Deutsche Forschungsgemeinschaft.


\appendix
\section{Free Bohmian particle in the plane}
\label{Bohmianfree}

\subsection{Time propagator}

The time propagator for a free quantum particle in the {$d-$dimensional} plane is given by, see e.g.~\cite{CohenTannoudji1966}:
\begin{equation}
  \label{prop_free}
  K({\bf x},t;{\bf x}_0,0)=\left(\dfrac{m}{2\pi\ic\hbar t}\right)^{d/2}\ee^{\ic m\frac{({\bf x}-{\bf x}_0)^2}{2\hbar t}}, t>0\ ,
\end{equation}

This explicit expression allows one to check easily one general property of the time propagators.
The quantum time propagator is not an element in the Hilbert space L$^2(\rR^d)$, hence, within standard mechanics, it cannot be considered as a physically observable state.

\subsection{Bohmian velocity field computed from the propagator}
The Bohmian velocity field corresponding to the propagator (\ref{prop_free}) is
\begin{equation}
 {\bf v}^\psi({\bf x},t)=\frac{{\bf x}-{\bf x}_0}{t},\ t>0\ {.}
\label{Bohmian_vfieldfree}
\end{equation}
So the trajectories are straight lines starting at ${\bf x}_0$:
\begin{equation}
 {\bf x} = {\bf x}_0 + {\bf c} t,\ t>0\ {,}
\label{Bohmian_vfieldfree}
\end{equation}
where ${\bf c}$ is a constant vector.

Note that the corresponding density
\begin{equation}
  |K({\bf x},t;{\bf x}_0,0)|^2=\left(\dfrac{m}{2\pi\hbar t}\right)^d
\end{equation}
is not normalisable when integrating over ${\bf x}$, which is another way to see that the propagator cannot be a physical state. This also means that it does not yield a probability measure over trajectories.

\subsection{Bohmian velocity field computed from a Gaussian wave packet}

The explicit expression of the quantum propagator (\ref{prop_free}) can be used to compute how a Gaussian wave packet propagates in the two$-$dimensional plane. As the problem is separable in Cartesian coordinates we will start to treat a one$-$dimensional wave packet. Consider the following initial Gaussian wave packet:
\begin{equation}
  \label{initGWP_1D}
  \psi_0(x)=\dfrac{\ee^{-\frac{(x-\overline{x})^2}{4\sigma^2}+\ic \frac{\overline{p} x}{\hbar}}}{(2\pi\sigma^2)^{1/4}}\ .
\end{equation}
It is centred at the position $\overline{x}$, {and} has an initial width $\sigma$ in position and an initial momentum $\overline{p}$. In order to determine how it is propagating, one can e.g. compute the convolution of {(\ref{initGWP_1D})} with the one$-$dimensional analogue of (\ref{prop_free}). 
\begin{equation}
  \psi(x,t)=\int_{-\infty}^\infty \sqrt{\dfrac{m}{2\pi\ic\hbar t}}\ee^{\ic m\frac{({x}-{y})^2}{2\hbar t}}\psi_0(y)\ud y\ .
\end{equation}
Doing the change of variable $u\equiv y-\overline{x}$ leads to a standard Gaussian integral and the result can be written:
\begin{equation}
  \label{GWP_1D}
  \psi(x,t)=\dfrac{\ee^{\ic\frac{ m (x-\overline{x})^2}{2 \hbar t} +\ic\frac{\overline{p}\,\overline{x}}{\hbar}}}{(2\pi\sigma^2)^{1/4}\sqrt{1+\frac{\ic \hbar t}{2m\sigma^2}}}
\exp\left\{ -\dfrac{\ic m}{2 \hbar t}\dfrac{(x-\overline{x}-\frac{\overline{p} t}{m})^2}{1+\frac{\ic \hbar t}{2m\sigma^2}} \right\}\ .
\end{equation}
In particular one recovers that the centre of the wave packet follows the trajectory:
\begin{equation}
  \label{GWP_1D_center}
  \langle x\rangle\equiv \int_{-\infty}^\infty x \left|\psi(x,t)\right|^2\ud x=\overline{x}+\frac{\overline{p} t}{m}\ ,
\end{equation}
and that the width of the wave packet is:
\begin{equation}
  \label{GWP_1D_width}
  \Delta x=\sqrt{\langle x^2\rangle-\langle x\rangle^2}=\sigma\left| 1+\frac{\ic \hbar t}{2m\sigma^2}\right|\equiv \sigma(t) \ .
\end{equation}
Using (\ref{defvpsi}) the Bohmian velocity field for a one$-$dimensional free Gaussian wave packet is:
\begin{equation}
  \label{Bohm_vfield_GWP_1D}
  v^\psi(x,t)=\dfrac{\dfrac{\overline{p}}{m}+\left(\dfrac{\hbar t}{2m\sigma^2}\right)^2 \dfrac{x-\overline{x}}{t}}{1+\left(\dfrac{\hbar t}{2m\sigma^2}\right)^2}\ .
\end{equation}
The corresponding trajectories are presented in \cite{holland93b}.

The probability density is:
\begin{equation}
  \label{GWP_1D_rho}
  |\psi(x,t)|^2=\dfrac{\ee^{-\frac{ (x-\overline{x}-\overline{p} t/m)^2}{2 \sigma( t)^2} }}{\sqrt{2\pi\sigma(t)^2}} \ . 
\end{equation}
The expression (\ref{Bohm_vfield_GWP_1D}) can be used to deduce the result in the two$-$dimensional case. If one assumes for simplicity that the initial Gaussian wave packet has the same width in the $x-$ and in the $y-$direction, {the Bohmian velocity field is:}
\begin{equation}
  \label{Bohm_vfield_GWP_2D}
  {\bf v}^\psi(x,t)=\dfrac{\dfrac{{\bf \overline{p}}}{m}+\left(\dfrac{\hbar t}{2m\sigma^2}\right)^2 \dfrac{{\bf x}-{\bf \overline{x}}}{t}}{1+\left(\dfrac{\hbar t}{2m\sigma^2}\right)^2}\ .
\end{equation}
In particular it is worth noticing that we recover the expression (\ref{Bohmian_vfieldfree}) at large distance and at long time.

\section{Further results for the quantum propagator of the scattering problem by a half-line barrier}

We detail some properties of the quantum time propagator \eqref{prop_halflineN}. First its large distance asymptotics is recalled. Second the corresponding Bohmian trajectories are displayed.

\subsection{Large distance asymptotics}
\label{farfield}
We investigate here the far-field asymptotics of the quantum time propagator \eqref{prop_halflineN} for our scattering problem, for both Neumann and Dirichlet boundary conditions. Using the asymptotic expansion of the function $F(u)$ in the propagator, we find that in the far field, the Bohmian velocity field becomes radial in all directions. 

\begin{figure}[!ht]
    \begin{center}
  \includegraphics[width=0.6\textwidth]{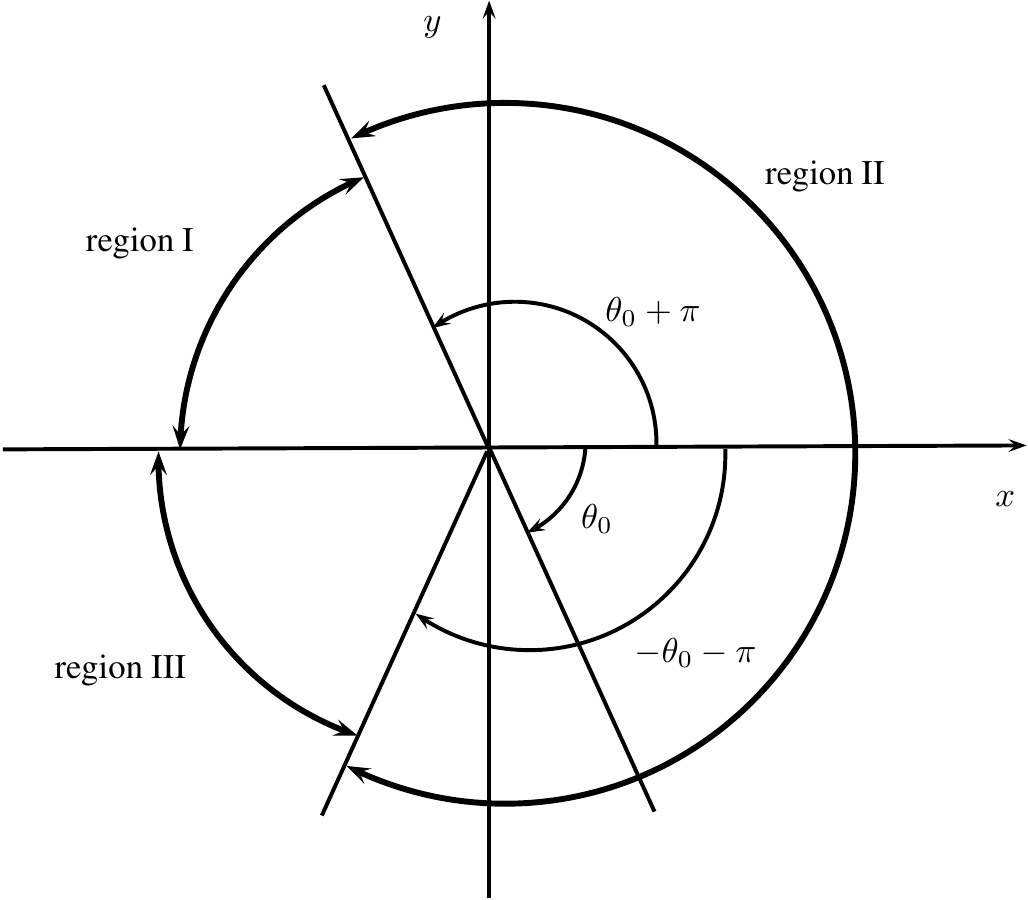}
  \end{center}
  \caption{Regions I, II and III, corresponding to different asymptotic behaviour. The angle $\theta_0$ is the polar angle of the source point.\label{regions}}   
\end{figure}

The asymptotic behaviour of $F(u)$ can be obtained by expressing it with more usual special functions. Indeed it is customary to identify $F(u)$ defined by (\ref{defF}) as related to Fresnel integrals via: 
\begin{equation}
F(u)=\ee^{-\ic u^2}\left[ \frac{1}{2}+\frac{\ee^{-\ic \pi/4}}{\sqrt{2}}\left( C\left(u\right)+\ic 
S\left(u\right) \right) \right]\ ,
\end{equation}
with the Fresnel integrals defined by, see e.g. \cite{gradshteyn07}:
\begin{equation}
  C(x)=\sqrt{\frac{2}{\pi}}\int_0^x \cos t^2\ud t,\qquad S(x)=\sqrt{\frac{2}{\pi}}\int_0^x \sin t^2\ud t\ .
\end{equation}
Therefore one can deduce its large argument asymptotics, see also \cite{lewis69,schulman82}:
\begin{eqnarray}
F(u)\simeq \ee^{-\ic u^2 }\left[ \Theta(u)-\dfrac{\ee^{\ic u^2+\ic\pi/4}}{2\sqrt{\pi}{u}}\right] \qquad {\textrm{for }}
u\to\pm\infty\ ,\label{Flarge}
\end{eqnarray}
where $\Theta(u)$ is the Heaviside step function. The asymptotic behaviour of the propagator \eqref{prop_halflineN} depends on the signs of $u_1$ and $u_2$. Taking (without loss of generality) $-\pi < \theta_0 \le 0$, we have 
\begin{equation}
  \label{optb1}
  -\pi \le \theta \le \pi+\theta_0 \Leftrightarrow {u_1 \ge 0}\ , \qquad
   -\pi -\theta_0 \le \theta < \pi \Leftrightarrow {u_2 \le 0}\ .
\end{equation}
As such, there are three regions with different asymptotic behaviour, see Fig.~\ref{regions}:
\begin{itemize}
\item Region I: $\pi+\theta_0 < \theta < \pi$, i.e., $u_1 < 0$ and $u_2 < 0$; the total wave is a scattered wave
\item Region II: $-\pi -\theta_0 < \theta < \pi+\theta_0$, i.e., {$u_1 > 0$ and $u_2 < 0$}; the total wave is the sum of the incident wave and a scattered wave
\item Region III: $-\pi < \theta < -\pi -\theta_0$, i.e., $u_1 > 0$ and $u_2 > 0$; the total wave is the sum of the incident wave and a reflected wave
\end{itemize}
There are two optical boundaries, i.e.\ two directions where the far field asymptotics significantly changes, given by $\theta_0 +\pi$ and ${-\theta_0 -\pi}$. 

\
In order to employ \eqref{Flarge}, we need to consider large $u_1$ and $u_2$. Hence ${\bf x}$ cannot be too close to one of the optical boundaries (where either $u_1=0$ or $u_2=0$). In addition, $\dfrac{m r r_0}{\hbar t}$ needs to be large. For region I, this leads to 
\begin{multline}
\label{asypmt1}
K_{N,D}({\bf x},t|{\bf x}_0,0)  \simeq\\
{\frac{1}{4\pi \ic}\sqrt{\frac{m}{2\pi \hbar t r r_0}} \ee^{\ic \frac{m(r+r_0)^2}{2\hbar t} + \ic \pi/4} \left[  -\frac{1}{\cos\left(\dfrac{\theta-\theta_0}{2}\right) }\mp \frac{1}{\cos\left(\dfrac{\theta+\theta_0}{2}\right) }  \right]} \ ,
\end{multline}
so that at leading order the Bohmian velocity field is:
\begin{equation}
{\bf v}_{N,D}\simeq \left(\frac{r+r_0}{t}\right)\frac{{\bf x}}{r},\quad 1 \ll r,r_0 \textrm{ and }\pi+\theta_0 \le \theta \le \pi\ .
\end{equation}

In region II, we have
\begin{equation}
  \label{asypmt2}
K_{N,D}({\bf x},t|{\bf x}_0,0){\simeq} \frac{m}{2\pi\ic \hbar t }\ee^{\ic \frac{m ({\bf x}- {\bf x_0})^2}{2\hbar t}} 
\end{equation}
with Bohmian velocity field
\begin{equation}
  {\bf v}_{N,D}\simeq \frac{{\bf x}-{\bf x_0}}{t},\quad 1 \ll r,r_0 \textrm{ and }-\pi -\theta_0 \le \theta \le \pi+\theta_0\ .
\end{equation}

Lastly, in region III, we have
\begin{equation}
  \label{asypmt3}
K_{N,D}({\bf x},t|{\bf x}_0,0){\simeq} \frac{m}{2\pi\ic \hbar t }\left[\ee^{\ic \frac{m ({\bf x}- {\bf x_0})^2}{2\hbar t}}
\pm \ee^{\ic \frac{m ({\bf x}- {\bf x'_0})^2}{2\hbar t}}\right] \ ,
\end{equation}
where
{\begin{equation}
{\bf x'_0}=(x_0,-y_0)\ , 
\label{x0prime}
\end{equation}
is the mirror image of ${\bf x_0}$ under reflection along the $x-$axis.} The Bohmian velocity field is:
\begin{equation}
  {\bf v}_{N,D}\simeq \re\left[\dfrac{\frac{{\bf x}-{\bf x_0}}{t} \ee^{\ic \frac{m ({\bf x}- {\bf x_0})^2}{2\hbar t}} \pm \frac{{\bf x}-{\bf x'_0}}{t} \ee^{\ic \frac{m ({\bf x}-{\bf x'_0})^2}{2\hbar t}} }{\ee^{\ic \frac{m ({\bf x}- {\bf x_0})^2}{2\hbar t}}
\pm \ee^{\ic \frac{m ({\bf x}- {\bf x'_0})^2}{2\hbar t}}} \right] = \frac{1}{t}({\bf x} - x_0 {\bf e}_x ) \ .
\end{equation}

In each case, the Bohmian velocity field becomes radial in the limit where $x\gg x_0$ and $y\gg y_0$.

\subsection{Bohmian trajectories computed from the propagator}
\label{Traj_quantum_prop}

\begin{figure}[!ht]
  \begin{minipage}[l]{0.49\linewidth}
    \begin{center}
      \includegraphics[width=\textwidth]{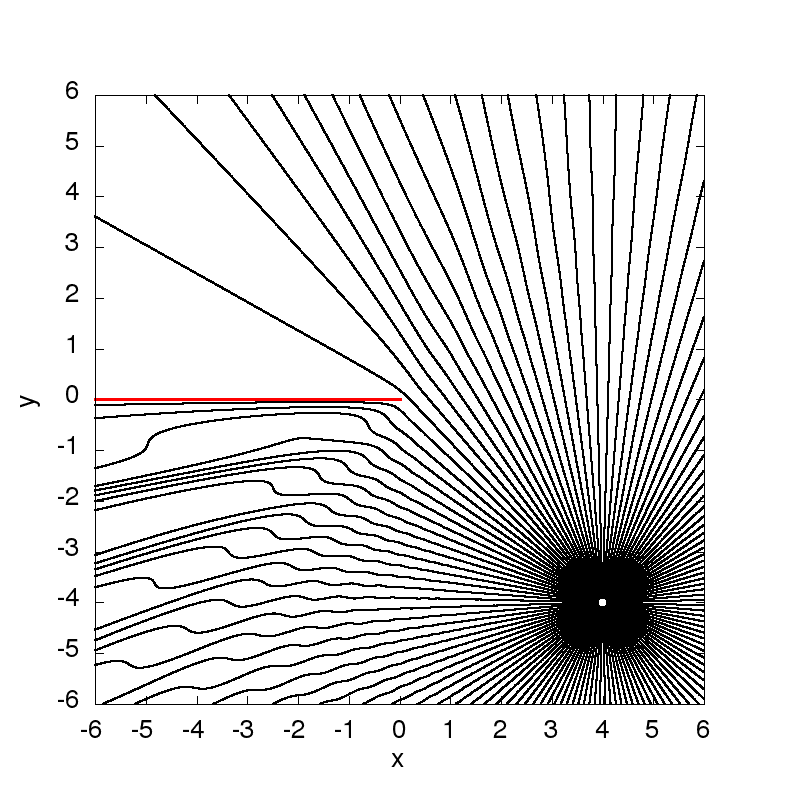}
    \end{center}
  \end{minipage}
  \begin{minipage}[r]{0.49\linewidth}
    \begin{center}
      \includegraphics[width=\textwidth]{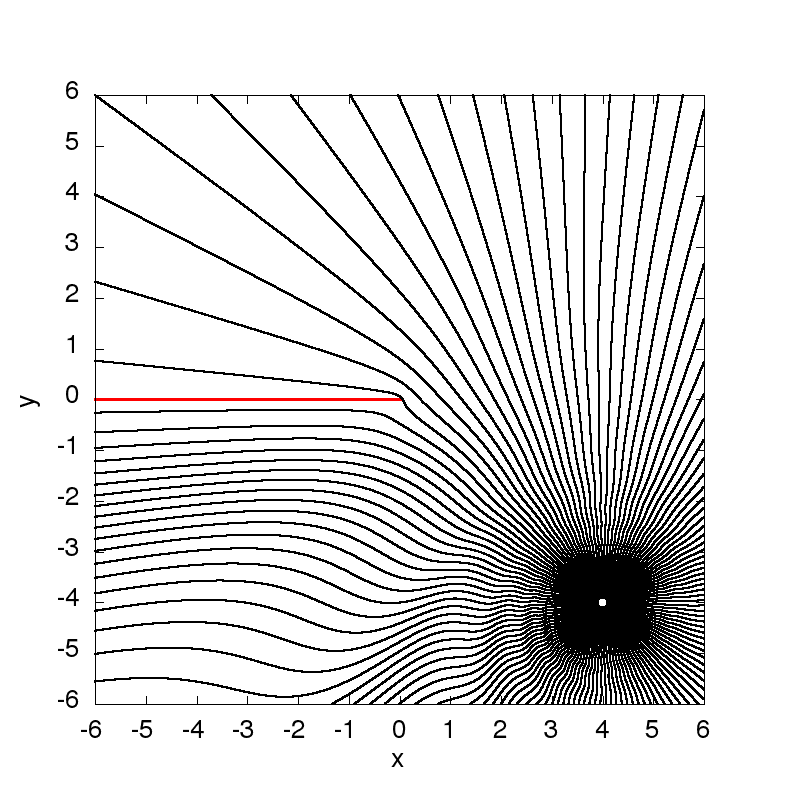}
    \end{center}
  \end{minipage}
  \caption{
Bohmian trajectories in the presence of a half-line obstacle with Neumann boundary conditions. 
The velocity field is computed {for} the propagator (\ref{prop_halflineN}), i.e.\ the ``initial state'' is a Dirac distribution supported at ${\bf x}_0=(4,-4)$.
The initial positions are along a circle centred at $(x_0,y_0)$ with radius $\rho=0.1$ at time $t_{\rm init}$. Left: {$t_{\rm init}=0.01$. Right: $t_{\rm init}=0.1$.}
}\label{Neumann_prop_x4_vs_tinit}
\end{figure}

\begin{figure}[!ht]
  \begin{minipage}[l]{0.49\linewidth}
    \begin{center}
      \includegraphics[width=\textwidth]{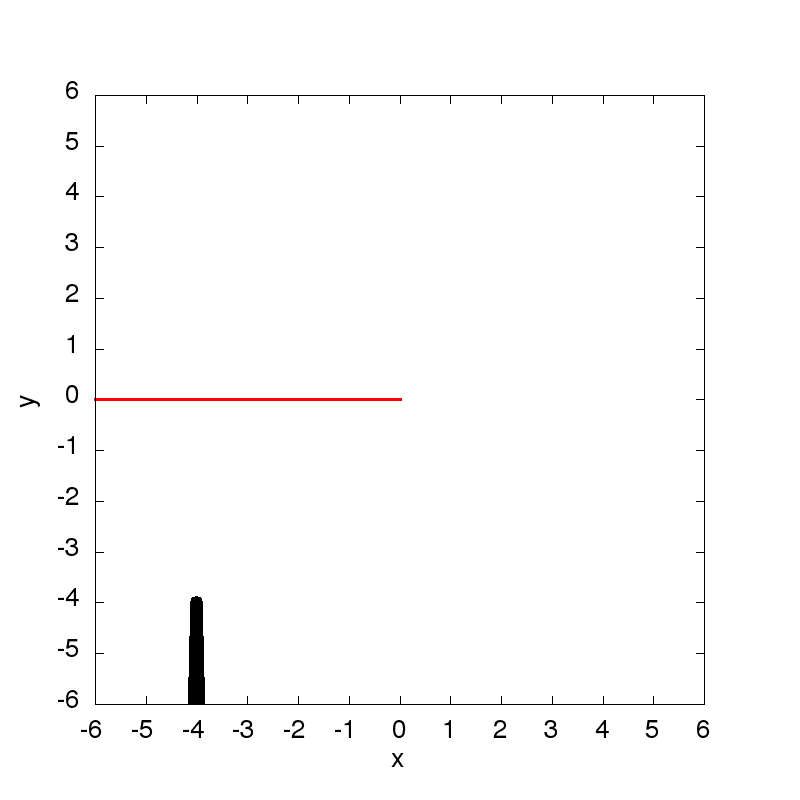}
    \end{center}
  \end{minipage}
  \begin{minipage}[r]{0.49\linewidth}
    \begin{center}
      \includegraphics[width=\textwidth]{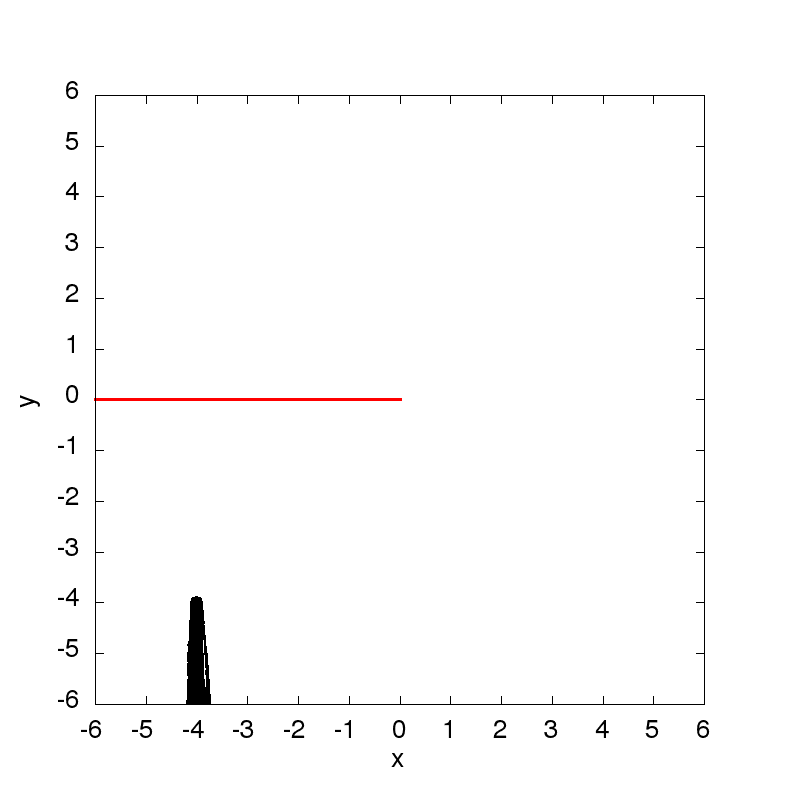}
    \end{center}
  \end{minipage}
  \caption{Same as in Fig.~\ref{Neumann_prop_x4_vs_tinit} for {$(x_0,y_0)=(-4,-4)$.}}
\label{Neumann_prop_x-4_vs_tinit}
\end{figure}

\begin{figure}[!ht]
   \begin{minipage}[l]{0.49\linewidth}
    \begin{center}
      \includegraphics[width=\textwidth]{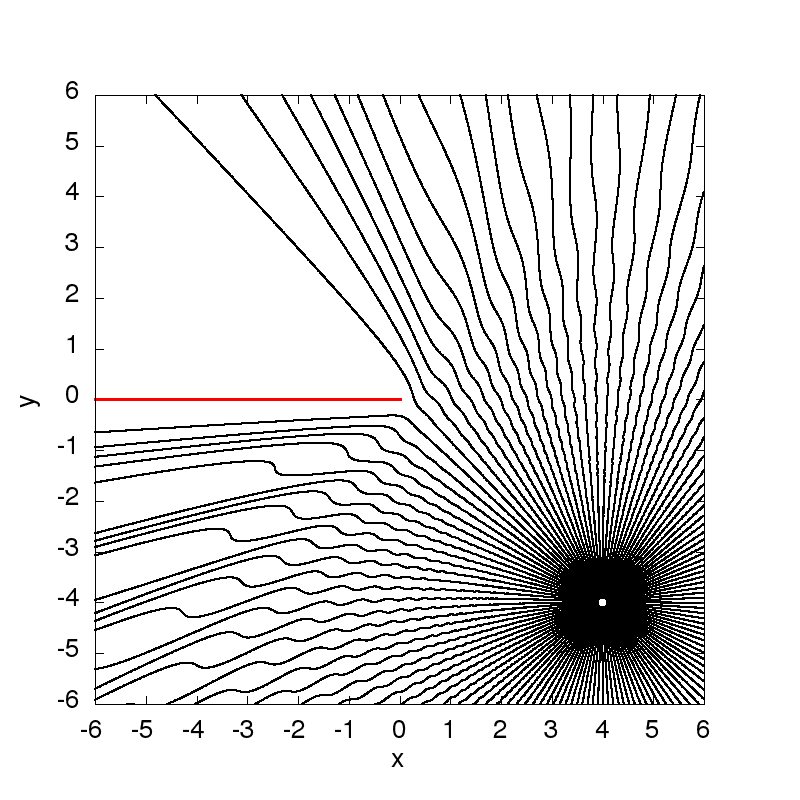}
    \end{center}
  \end{minipage}
  \begin{minipage}[r]{0.49\linewidth}
    \begin{center}
      \includegraphics[width=\textwidth]{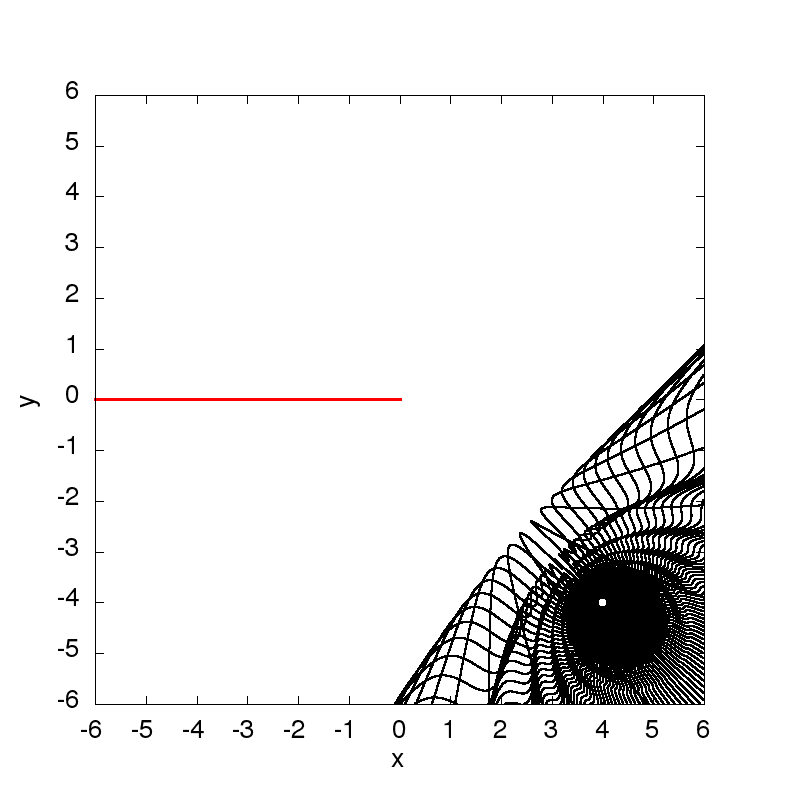}
    \end{center}
  \end{minipage}
  \caption{
Bohmian trajectories in the presence of a half-line obstacle with Dirichlet boundary conditions. 
The velocity field is computed {for} the propagator (\ref{prop_halflineN}), i.e.\ the ``initial state'' is a Dirac distribution supported at ${\bf x}_0=(4,-4)$.
The initial positions are along a circle centred at $(x_0,y_0)$ with radius $\rho=0.1$ at time $t_{\rm init}$. Left:{ $t_{\rm init}=0.01$. Right: $t_{\rm init}=0.1$.}
}
  \label{Dirichlet_prop_x4_vs_tinit}
 \end{figure}

\begin{figure}[!ht]
   \begin{minipage}[l]{0.49\linewidth}
    \begin{center}
      \includegraphics[width=\textwidth]{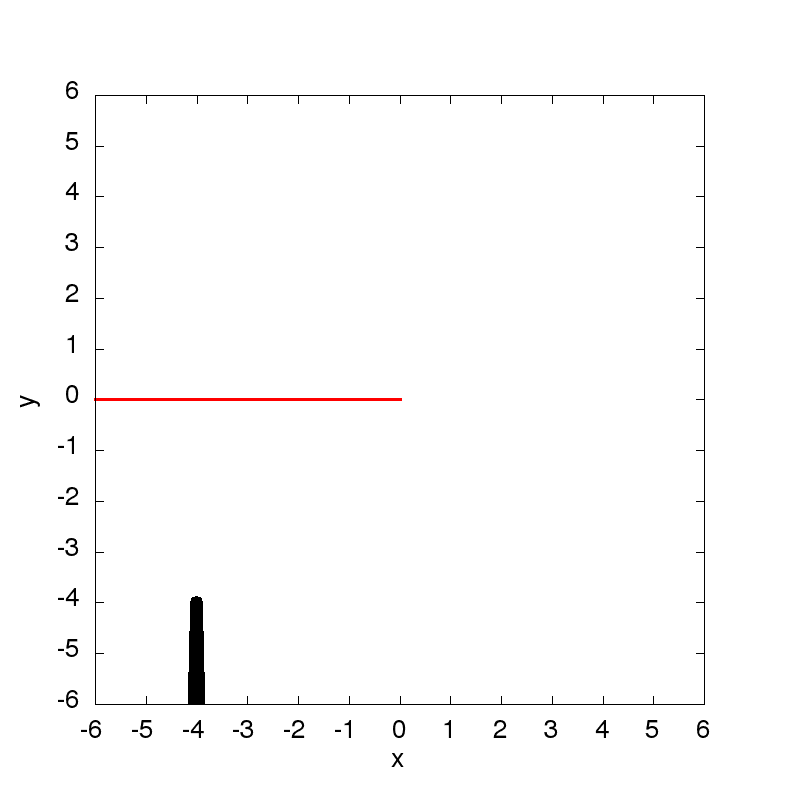}
    \end{center}
  \end{minipage}
  \begin{minipage}[r]{0.49\linewidth}
    \begin{center}
      \includegraphics[width=\textwidth]{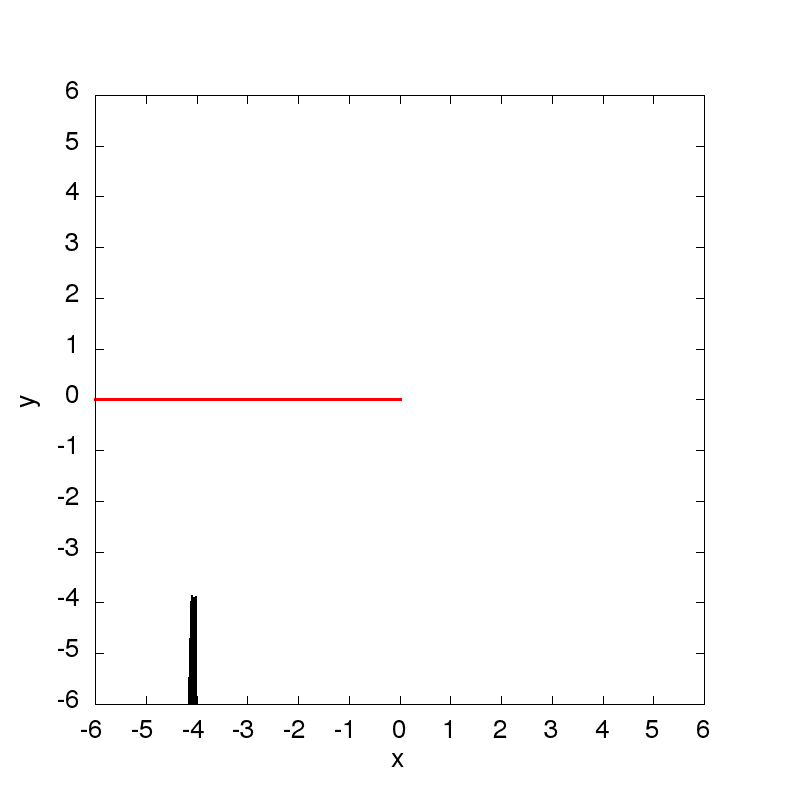}
    \end{center}
  \end{minipage}
  \caption{Same as in Fig.~\ref{Dirichlet_prop_x4_vs_tinit} for {$(x_0,y_0)=(-4,-4)$.}}   \label{Dirichlet_prop_x-4_vs_tinit}
\end{figure}

The gradient of the propagator (\ref{prop_halflineN}) is given by 
\begin{eqnarray}
\hspace{-1cm}
{\boldsymbol \nabla} K_{N,D} &=& \frac{m^2}{2\pi\ic\hbar^2 t^2} \ee^{\ic \frac{m(r+r_0)^2}{2\hbar t}} {\Bigg\{ \ic \left[({\bf x} -{\bf x}_0)F(u_1) \pm ({\bf x} -{\bf x}'_0)F(u_2)  \right]} \nonumber\\ 
&& + \sqrt{\dfrac{r_0 \hbar t}{2\pi m r}} \ee^{-\ic \pi/4}  \Bigg[ \left( \cos \left(\frac{\theta + \theta_0}{2}\right) \mp \cos \left(\frac{\theta - \theta_0}{2}\right) \right) {\bf e}_x \nonumber\\ 
&& + {\left( \sin \left(\frac{\theta + \theta_0}{2}\right) \mp \sin \left(\frac{\theta - \theta_0}{2}\right)  \right) {\bf e}_y} \Bigg] \Bigg\}\ ,
\label{dK}
\end{eqnarray}
where {${\bf x'_0}$ is defined in \eqref{x0prime}.} This yields the Bohmian velocity field
\begin{eqnarray}
\hspace{-1cm}
 {\bf v}_{N,D} &=& \frac{1}{t} \Bigg\{  {\bf x} - x_0 {\bf e}_x - y_0 {\bf e}_y \frac{{|F(u_1)|^2 - |F(u_2)|^2}}{|F(u_1) \pm F(u_2)|^2}  \nonumber\\ 
&& + \sqrt{\dfrac{r_0 \hbar t}{2\pi m r}}  {\textrm{Im}}\left(\frac{\ee^{-\ic \pi/4}}{F(u_1) \pm F(u_2)}\right) \Bigg[ \left( \cos \left(\frac{\theta + \theta_0}{2}\right) \mp \cos \left(\frac{\theta - \theta_0}{2}\right) \right) {\bf e}_x \nonumber\\ 
&& +  \left(  \sin \left(\frac{\theta + \theta_0}{2}\right) \mp \sin \left(\frac{\theta - \theta_0}{2}\right)  \right) {\bf e}_y \Bigg] \Bigg\}
\label{velocity}
\end{eqnarray}
The Bohmian trajectories can easily be computed using these expressions and 4th order Runge Kutta method to solve (\ref{Bohm_eq}). They are displayed in Figs.~\ref{Neumann_prop_x4_vs_tinit} and \ref{Neumann_prop_x-4_vs_tinit} for Neumann boundary conditions on the barrier, in Figs.~\ref{Dirichlet_prop_x4_vs_tinit} and \ref{Dirichlet_prop_x-4_vs_tinit} for Dirichlet boundary conditions on the barrier. In Figs.~\ref{Neumann_prop_x4_vs_tinit} and \ref{Dirichlet_prop_x4_vs_tinit}, one can see that the trajectories follow an intuitive radial behaviour around the initial point at small time. Conversely, in Figs.~\ref{Neumann_prop_x-4_vs_tinit} and \ref{Dirichlet_prop_x-4_vs_tinit}, the Bohmian particle is immediately repelled as the initial wave immediately interferes with the reflected wave.

\section{Bohmian trajectories for an incoming plane wave with Dirichlet boundary conditions on the barrier}
\label{pw_D}

\begin{figure}[!ht]
  \begin{minipage}[l]{0.49\linewidth}
    \begin{center}
      \includegraphics[width=\textwidth]{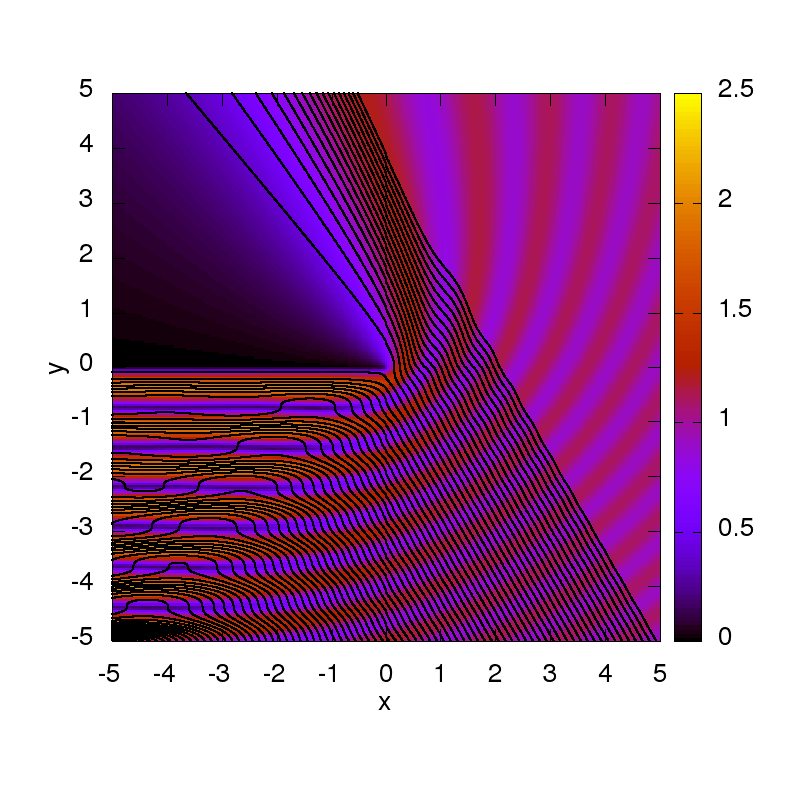}
    \end{center}
  \end{minipage}
  \begin{minipage}[r]{0.49\linewidth}
    \begin{center}
      \includegraphics[width=\textwidth]{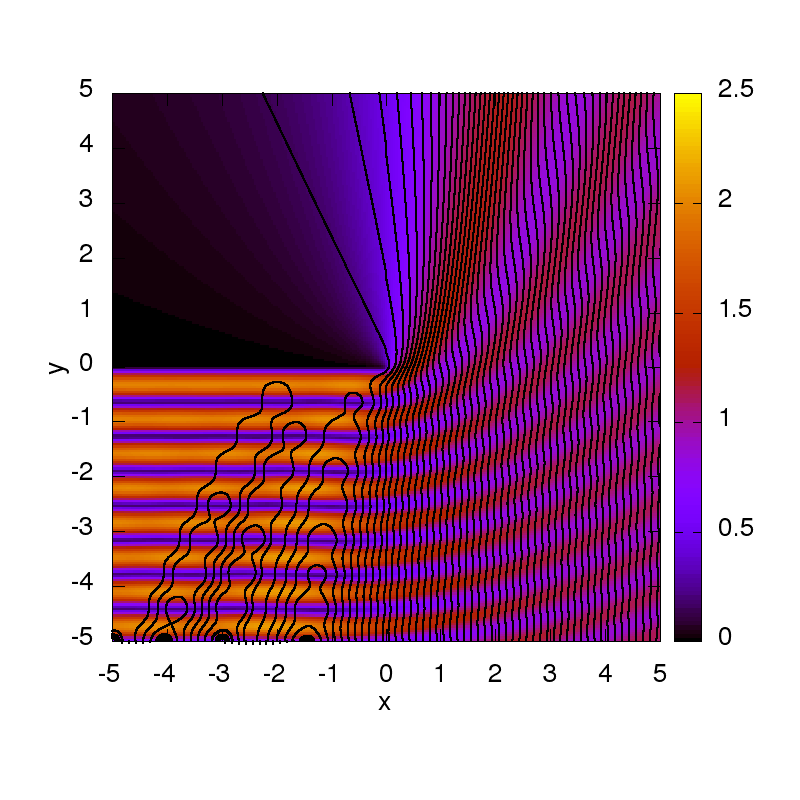}
    \end{center}
  \end{minipage}
  \caption{Bohmian trajectories in the presence of a half-line obstacle with Dirichlet boundary conditions. The velocity field is computed {for} an incident plane wave (\ref{scat_stateD}) with $k_0=5$. The background colour map stands for the modulus of the scattering state (\ref{scat_stateD}). Left: $\theta_0=\pi/3$. Right: $\theta_0=\pi/2$.}
\label{Bohm_traj_pwD}
\end{figure}

For sake of completeness the Bohmian trajectories have {also} been computed for an incoming plane wave with Dirichlet boundary conditions in the half-line barrier. The wave function is, similarly to (\ref{scat_state}), given by
\begin{equation}
   \psi({\bf x})=\ee^{\ic k_0 r}\left[F\left(a_1\right) - F\left(a_2\right)\right]\ ,\label{scat_stateD}
\end{equation}
with the same definition as before. Bohmian trajectories can be computed following the same steps as for Neumann boundary conditions. Therefore we only give here the formula for the gradient 
\begin{equation}
   {\boldsymbol \nabla} \psi = \ee^{\ic k_0 r} \Bigg\{ \ic \left[F(a_1) {\bf k_0}- F(a_2) {\bf k'_0}\right] + \ee^{-\ic\pi/4} \sqrt{\frac{2k_0}{\pi r}} \cos\left(\frac{\theta_0}{2}\right) \left[ \cos\left(\frac{\theta}{2}\right){\bf e}_x + \sin\left(\frac{\theta}{2}\right){\bf e}_y \right] \Bigg\} \ ,
\end{equation}
and the Bohmian velocity field
\begin{multline}
{\bf v}^\psi = \frac{\hbar k_x }{m}  {\bf e}_x + \frac{\hbar  k_y}{m} {\bf e}_y \frac{|F(a_1)|^2 -|F(a_2)|^2 }{|F(a_1)-F(a_2)|^2} \\
+ \frac{\hbar}{m}\sqrt{\frac{2k_0}{\pi r}} {\textrm{Im}}\left( \frac{\ee^{-\ic\pi/4}}{F(a_1)-F(a_2)} \right)\cos\left(\frac{\theta_0}{2}\right) \left[ \cos\left(\frac{\theta}{2}\right){\bf e}_x + \sin\left(\frac{\theta}{2}\right){\bf e}_y \right]  \ .
\end{multline}
The trajectories are displayed in Fig.~\ref{Bohm_traj_pwD}. {The trajectories are very similar to the ones in the case of Neumann boundary conditions.}
\pagebreak


\end{document}